\documentclass[twocolumn]{aastex7}

\usepackage{amsmath,amssymb}
\usepackage{graphicx}
\usepackage{bm}
\usepackage{subcaption}

\shorttitle{Sub-mm GW Counterparts with XSMT}
\shortauthors{Xiao}

\begin{document}
	
	\title{Submillimeter Detectability of Gravitational-Wave Counterparts from Neutron-Star Mergers with the Xue-shan-mu-chang 15-meter Telescope}

	\author[orcid=0000-0002-4304-2759,sname='Xiao']{Di Xiao}
	\affiliation{
		Purple Mountain Observatory, Chinese Academy of Sciences, Nanjing 210023, China; dxiao@pmo.ac.cn
	}
	\affiliation{
		State Key Laboratory of Radio Astronomy and Technology, Purple Mountain Observatory, Chinese Academy of Sciences, 10 Yuanhua Road, Nanjing 210023, China
	}
	\email{dxiao@pmo.ac.cn}
	
	\begin{abstract}

        Submillimeter (sub-mm) follow-up of binary neutron star (BNS) mergers provides unique constraints on the early-time energetics and environments of relativistic outflows, capturing the spectral evolution at epochs where centimeter-band emission is often still optically thick or yet to peak. However, the practical scientific yield depends on instrument-specific thresholds, the observing cadence, and the distinct temporal contributions from isotropic ejecta versus beamed relativistic jets. With the upcoming Xue-shan-mu-chang 15-meter SubMillimeter Telescope (XSMT), facility-specific forecasts are needed to test for sustained engine energy injection, as expected for a long-lived magnetar remnant rather than a promptly formed black hole. We present a unified numerical framework that couples engine-driven ejecta dynamics to non-thermal synchrotron emission, accounting for synchrotron self-absorption and deep-Newtonian effects. Adopting fixed 5$\sigma$ (1 h) point-source thresholds of 1.5/2.9/10.2 mJy at 230/345/460 GHz, we construct parameter-space detectability maps and estimate event rates based on current BNS merger-rate priors. For a fiducial local event at 40 Mpc, we find that a magnetar-boosted ejecta afterglow peaks on timescales of weeks to months and remains detectable long enough to allow delayed follow-up, with an expected all-sky rate of $\dot N_{\rm ej} \approx 0.05$--1.7 yr$^{-1}$ at 230 GHz for $f_{\rm mag}=1$; this rate is an upper limit and scales linearly with the long-lived magnetar fraction. Conversely, while relativistic jets produce intense early-time signals, their detection is constrained by narrow beaming and fleeting visibility. These results motivate a two-tier Target-of-Opportunity strategy: rapid-response visits to capture early jet emission, and sparse long-baseline monitoring optimized for the persistent ejecta afterglow. Our framework provides a quantitative basis for prioritizing gravitational-wave triggers and maximizing the scientific yield of XSMT in the multi-messenger era.
	\end{abstract}
	
	
	\section{Introduction}

    The detection of gravitational waves (GW) from the binary neutron star (BNS) merger GW170817 \citep{Abbott2017a}, followed by the identification of multi-wavelength electromagnetic (EM) counterparts \citep{Abbott2017b,Chornock2017,Goldstein2017,Haggard2017,Hallinan2017,Margutti2017,Savchenko2017,Troja2017,ZhangBB2018}, has transformed our understanding of relativistic outflows and cosmic nucleosynthesis. In the wake of this landmark event, the focus of the astrophysical community has shifted toward characterizing the nature of the merger remnant—specifically, distinguishing between the prompt formation of a black hole (BH) and the birth of a long-lived, highly magnetized millisecond magnetar \citep{Piro2017,Ai2018,Sarin2021,Plasse2026}. While optical and near-infrared (NIR) observations of kilonovae have confirmed the r-process origin of heavy elements \citep{Li1998,Kulkarni2005,Metzger2010,Kasen2013,Andreoni2017,Arcavi2017,Coulter2017,Drout2017,Kilpatrick2017,Lipunov2017,Pian2017,Shappee2017,Smartt2017,Soares-Santos2017,Utsumi2017,Valenti2017,Villar2017}, the sub-millimeter (sub-mm) band remains a relatively unexplored window \citep{Alexander2017,Kim2017}. Occupying the frequency regime between the thermal peaks of the kilonova and the non-thermal radio afterglow \citep{Alexander2017,Margutti2018}, sub-mm follow-up provides complementary constraints on the energetics and environments of BNS mergers that are difficult to uniquely resolve at cm bands alone \citep{Metzger2014a,Hotokezaka2016}, particularly during the early weeks when centimeter emission is heavily suppressed by synchrotron self-absorption \citep[SSA; e.g.,][]{Sari1998,Granot2002,Hallinan2017,Sadeh2024}.

    The outcome of a BNS merger is primarily dictated by the total mass relative to the maximum mass supported by the nuclear equation of state \citep{Just2015,Ai2020,Sarin2021}. Numerical relativity simulations suggest that a non-negligible fraction of mergers produce a millisecond magnetar, capable of storing an immense rotational energy reservoir ($E_{\rm rot} \approx 2 \times 10^{52}$ erg for a 1 ms period). The generation of magnetar-strength fields ($\sim 10^{15}$--$10^{16}$ G) is thought to occur within milliseconds of the merger, driven by Kelvin-Helmholtz instabilities at the shear interface of the colliding stars \citep{Rasio1999,Price2006,Kiuchi2014,Giacomazzo2015, Kiuchi2015,Ciolfi2017,Shibata2021a,Shibata2021b,Combi2023,AguileraMiret2024,Kiuchi2024}. This massive energy injection profoundly alters the debris dynamics; whereas a BH remnant provides only $\sim 10^{49}$ erg via radioactive decay \citep{Metzger2010,Korobkin2012,Tanaka2013,Rosswog2014}, a magnetar engine can continuously power the ejecta through a magnetized wind, leading to an ``engine-fed kilonova" or ``mergernova" \citep{Yu2013,Metzger2014b,Gao2015,Kisaka2015,Metzger2017,Ai2022,Sarin2022,Omand2024,Ai2025}. While the choice of central engine dictates the energy budget, the dynamical and disk-wind processes during the merger supply the material substrate, expelling $10^{-4}$ to $10^{-2} M_\odot$ of neutron-rich ejecta \citep{Hotokezaka2013,Radice2016,Radice2018,Kawaguchi2022,Fujibayashi2023}. The expansion of this ejecta—potentially re-accelerated by the magnetar wind—eventually shocks the interstellar medium (ISM), driving a late-time afterglow \citep{Nakar2011,Piran2013,Margalit2015,Hotokezaka2015,Hotokezaka2018,Murase2018,Margalit2020,Ghosh2022,Merfeld2025}. Deep radio searches for this signal have been performed for GW170817 \citep{Hallinan2017,Dobie2018,Mooley2018,Balasubramanian2022,Katira2025} and a suite of historical short gamma-ray bursts  \citep{Fong2016,Schroeder2020,Ghosh2024}, providing critical constraints on the kinetic energy and ambient density of these events \citep{Liu2020}.

    In the sub-mm regime, the emission is characterized by a complex interplay between thermal and non-thermal components. While the band samples the Rayleigh-Jeans tail of the thermal kilonova (or ``mergernova" in engine-fed cases), this contribution is generally sub-dominant compared to non-thermal signatures at these frequencies. In this work, we focus on the two primary non-thermal channels: (i) the early-time synchrotron emission from a relativistic jet, which often undergoes an optically thick-to-thin transition in the sub-mm band, and (ii) the late-time ejecta afterglow produced as the bulk debris shocks the interstellar medium. The latter can be significantly boosted by energy injection from a magnetar wind, potentially explaining late-time X-ray excesses as seen in GW170817 \citep{Hajela2022}. By focusing on these non-thermal components, we address the most promising signals for discovery and characterization with upcoming facilities.
    
    Despite its potential, sub-mm follow-up is not simply ``radio at higher frequency". Detectability and optimal cadence depend sensitively on the distinct temporal footprints of these components: a relativistic jet may produce a bright early-time signal within hours, while the isotropic ejecta afterglow, modeled below as a spherical $\Omega=4\pi$ outflow, evolves over weeks to months and can remain visible over a much broader window \citep{Margalit2015,Lee2018,Gottlieb2019,Margalit2020}. For off-axis observers, the sub-mm peak often precedes the radio peak, providing an earlier diagnostic of the circumburst density ($n_{\rm ext}$) and microphysical parameters. Ultimately, the diverse nature of these emission components motivates a two-tier Target-of-Opportunity (ToO) strategy to capture both early-time bursts and late-time evolution.
    
    The forthcoming Xue-shan-mu-chang 15-meter SubMillimeter Telescope (XSMT) will provide continuum capability at 230/345/460 GHz \citep{XSMT2025}, enabling systematic sub-mm follow-up of nearby GW events. To bridge the gap between theoretical modeling and practical observation, we require a framework that connects physical uncertainties—such as energy injection efficiency and magnetic field amplification—to actionable observing plans. 
    
    Here we provide a facility-specific forecast for XSMT. We compute sub-mm light curves using a unified framework that couples engine-dependent ejecta dynamics (magnetar versus BH) to synchrotron emission including SSA and deep-Newtonian (DN) effects \citep{Huang2003,Granot2006,Sironi2013,Liu2020,MinhajurRahaman2026}. Here DN denotes the regime reached when the shock becomes sufficiently slow that the minimum electron Lorentz factor approaches unity and only a reduced fraction of electrons remains effectively radiating. We include a jet afterglow module with equal-arrival-time surface (EATS) corrections. We construct parameter-space detectability maps and estimate event rates based on current BNS merger-rate priors. Finally, we present a prioritized observing strategy designed to exploit the distinct temporal windows of the jet and ejecta channels.

    This paper is structured as follows. Section~\ref{sec:model} describes the physical model for the ejecta and jet afterglows. Section~\ref{sec:xsmt_sens} summarizes the adopted XSMT thresholds and detectability metrics. Section~\ref{sec:fiducial} presents fiducial light-curve predictions for a local (40~Mpc) merger. Section~\ref{sec:survey} provides parameter-space surveys and detectability maps. Section~\ref{sec:rates} derives detection-rate estimates, while Section~\ref{sec:strategy} proposes an observing strategy for XSMT. Section~\ref{sec:discussion} places the forecasts in context by comparing with previous studies and existing facilities, discussing key uncertainties and band prioritization, and addressing multiwavelength synergies, source confusion, and operational limitations. Our conclusions are summarized in Section~\ref{sec:conclusions}.
	
	\section{Physical Model}
	\label{sec:model}
	
	We describe the theoretical framework used to predict the sub-mm signatures of neutron-star mergers. We treat the observed emission as the sum of two distinct parts: (i) an isotropic afterglow produced by the merger ejecta and (ii) an optional relativistic jet afterglow. The behavior of the first component depends on the energy output from the central engine—either a long-lived magnetar or a BH—while the latter represents the beamed emission commonly linked to short gamma-ray bursts. The underlying dynamics and radiation details are documented in Appendices \ref{app:shared_shock}--\ref{app:ssa_dn}.
	
\subsection{Modeling Workflow Overview}
\label{subsec:workflow}

\begin{figure*}[t]
    \centering
    \includegraphics[width=\textwidth]{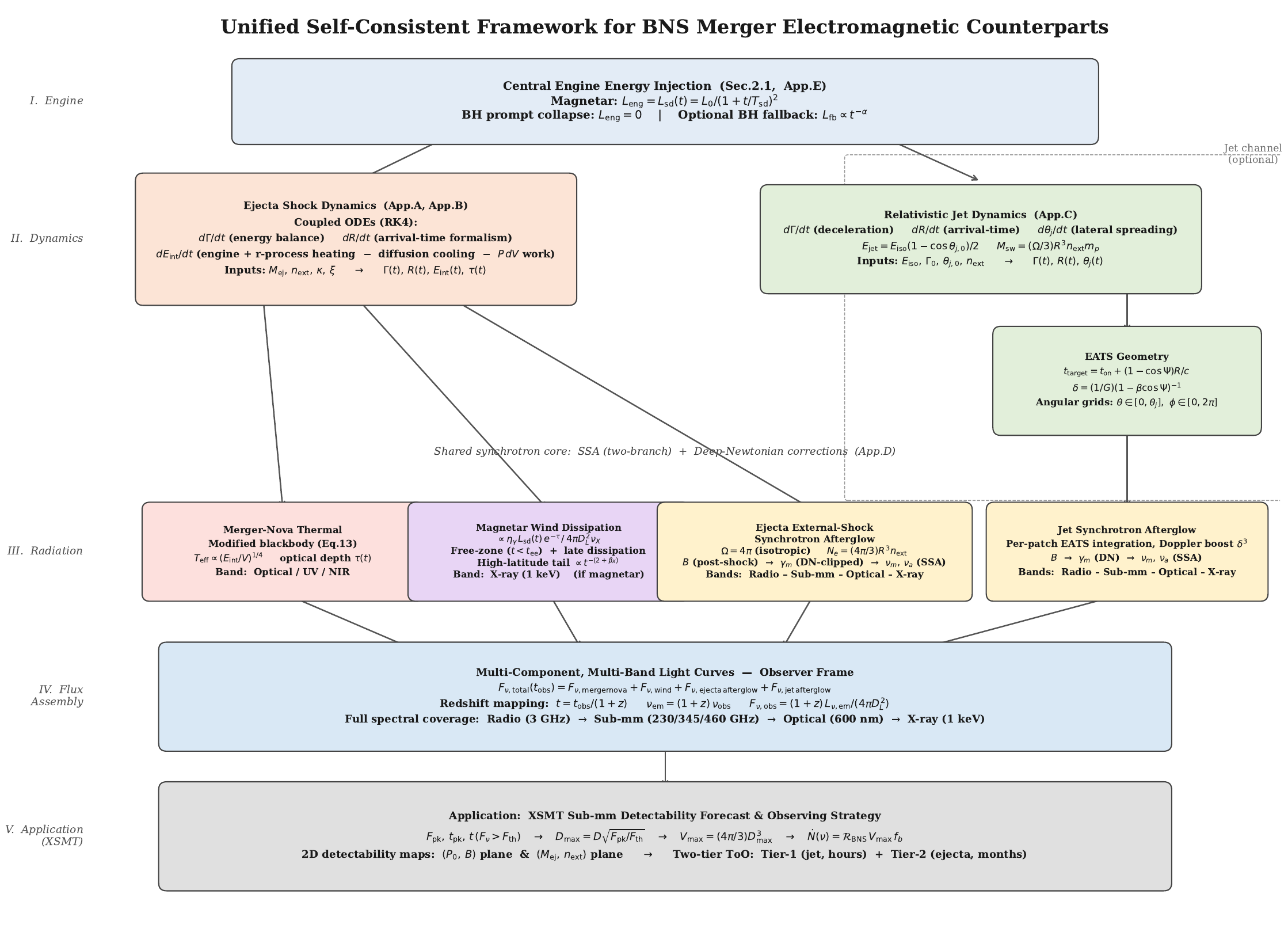}
    \caption{Schematic overview of the unified modeling framework. The pipeline proceeds through five stages: (I)~central-engine energy injection; (II)~ejecta and jet shock dynamics; (III)~multi-channel radiation (merger-nova thermal, magnetar wind X-ray, ejecta synchrotron afterglow, and jet synchrotron afterglow); (IV)~flux assembly and redshift mapping to the observer frame; and (V)~application to facility-specific detectability forecasts. The jet channel (right, dashed enclosure) is optional.}
    \label{fig:flowchart}
\end{figure*}

Figure~\ref{fig:flowchart} provides a schematic overview of the full computational pipeline. The calculation begins with the central engine, which sets the time-dependent power supplied to the merger ejecta: a long-lived magnetar injects dipole spin-down luminosity, whereas a prompt-collapse BH is represented by the no-injection limit $L_{\rm eng}=0$ (Section~\ref{subsec:engine}; Appendix~\ref{app:bh_engine}). This engine history is then coupled to the ejecta dynamics, giving the time evolution of the shock radius, bulk Lorentz factor, and internal energy (Sections~\ref{subsec:shared_shock} and \ref{subsec:ejecta_dyn}; Appendix~\ref{app:ejecta_dyn}). In parallel, when the jet module is enabled, the relativistic jet is evolved using the same arrival-time convention, with optional lateral spreading and equal-arrival-time-surface integration (Section~\ref{subsec:jet_model}; Appendix~\ref{app:jet}).

The dynamical outputs feed several radiation modules. The code retains thermal merger-nova and magnetar-wind dissipation channels for completeness, but the XSMT forecasts in this paper focus on the two non-thermal synchrotron components most relevant at sub-mm frequencies: the isotropic ejecta afterglow and the optional jet afterglow. Both synchrotron channels use the same external-shock radiation core, including SSA and DN corrections (Section~\ref{subsec:ejecta_afterglow}; Appendix~\ref{app:ssa_dn}). The resulting observer-frame light curves are then compared with the adopted XSMT thresholds (Section~\ref{sec:xsmt_sens}) to construct detectability maps, horizon distances, rate estimates, and observing strategies.

Our implementation is designed to connect, rather than replace, two earlier lines of modeling. \citet{Gao2015} developed a magnetar-powered merger framework emphasizing early X-ray/optical emission, including merger-nova radiation, magnetar-wind dissipation, and jet afterglows. \citet{Liu2020} focused on the late-time ejecta external-shock afterglow and its radio constraints. The present work combines these ingredients in a unified framework in which the engine luminosity is continuously coupled to the ejecta acceleration and the same dynamical backbone is used to generate multi-band afterglow predictions. This unified treatment allows us to make a direct, facility-specific forecast for XSMT at 230/345/460~GHz, rather than treating the early engine-powered emission, late ejecta afterglow, and observing strategy as separate problems.

\subsection{Central Engine and Energy Input}
	\label{subsec:engine}
	
	The choice of central engine primarily sets the total energy budget and the characteristic energy-injection timescale for the merger ejecta. If the merger remnant does not immediately collapse, we adopt a magnetar model where energy is supplied through dipole spin-down:
		\begin{equation}
			L_{\rm sd}(t)=\frac{L_0}{\left(1+t/T_{\rm sd}\right)^2},
			\label{eq:Lsd}
		\end{equation}
	where the initial power $L_0$ and the spin-down timescale $T_{\rm sd}$ are determined by the surface dipole magnetic field strength $B$ at the pole and the initial rotation period $P_0$: 
		\begin{align}
		L_0 &= \frac{B^2 R_{\rm ns}^6 \Omega_0^4}{6c^3}, \label{eq:L0}\\
		T_{\rm sd} &= \frac{3c^3 I}{B^2 R_{\rm ns}^6 \Omega_0^2} , \label{eq:Tsd}
	\end{align}
	and the initial rotational energy is $E_{\rm rot}=\tfrac12 I\Omega_0^2$ with $\Omega_0=2\pi/P_0$. For comparison, we also consider a BH engine where the remnant collapses promptly. In this scenario, we assume there is no long-term energy input ($L_{\rm eng}(t)=0$), meaning the ejecta evolves based only on its initial speed and internal heating from radioactive decay.  We optionally allow a phenomenological fallback power law for exploratory comparisons (Appendix~\ref{app:bh_engine}), but it is not required for the main results.
	
	\subsection{Shock Kinematics}
	\label{subsec:shared_shock}
	
	To keep the jet and ejecta components physically consistent, we evolve both using a unified time system based on when the light reaches the observer. We determine the shock radius $R(t)$ by integrating the growth rate in terms of the source-frame arrival time $t$, where $t = t_{\rm obs}/(1+z)$. The zero point of this clock is the onset of the central-engine/ejecta evolution. For the BNS remnants considered here, we take this onset to be coincident with the merger/GW trigger to within a negligible delay compared with the day-to-year timescales of the sub-mm afterglow. This approach is necessary for correctly calculating the emission from different parts of the expanding surface at the same observer time (Appendix \ref{app:shared_shock}). The motion is governed by how fast the shock moves and how quickly it sweeps up the surrounding matter, ensuring that energy and momentum are conserved as the flow slows down from relativistic speeds. 
	
	\subsection{Ejecta Evolution and Energy Injection}
\label{subsec:ejecta_dyn}

We model the merger ejecta as a shell of mass $M_{\rm ej}$ expanding into a circumburst medium of number density $n_{\rm ext}$. The engine prescription defined in Section~\ref{subsec:engine} enters the ejecta dynamics through the injected power $\xi L_{\rm eng}(t)$, where $\xi$ is the fraction of the engine luminosity that thermalizes in the ejecta. Thus the magnetar case inherits the dipole spin-down luminosity $L_{\rm sd}(t)$ from Eqs.~(\ref{eq:Lsd})--(\ref{eq:Tsd}), whereas the prompt-collapse BH case corresponds to the no-injection limit $L_{\rm eng}=0$.

The ejecta dynamics are governed by a coupled system of ordinary differential equations that track five variables as functions of the source-frame arrival time $t$: the bulk Lorentz factor $\Gamma(t)$, the shock radius $R(t)$, the comoving internal energy $E'_{\rm int}(t)$, the comoving volume $V'(t)$, and the comoving proper time $t'(t)$. To make the role of the engine explicit, we write the two energy equations that feed the bulk acceleration.

The motion of the ejecta, represented by the bulk Lorentz factor $\Gamma$, is determined by the overall energy balance of the system. In the laboratory frame, the total energy source or sink rate is given by \citep{Yu2013,Gao2015}:
\begin{align}
\frac{dE}{dt} &= \xi L_{\rm eng}(t)+\mathcal{D}^{2}L'_{\rm ra}-\mathcal{D}^{2}L'_{e},
\label{eq:dEdt_main}
\end{align}
where $\mathcal{D}=\Gamma(1+\beta)$ is the Doppler factor, $L'_{\rm ra}$ is the radioactive heating rate \citep{Korobkin2012}, $L'_e$ is the diffusion luminosity  (Appendix~\ref{app:ejecta_dyn}). The internal energy of the shell changes due to three main factors: energy input from the central engine, heating from the decay of heavy elements (r-process), and losses from expansion and radiation. These terms are collected in the comoving-frame evolution equation:
\begin{align}
\frac{dE'_{\rm int}}{dt'} &= \frac{\xi L_{\rm eng}(t)}{\mathcal{D}^{2}}+L'_{\rm ra}-L'_{e}-P'\frac{dV'}{dt'},
\label{eq:dEint_main}
\end{align}
 where $P'=E'_{\rm int}/(3V')$ is the comoving pressure.

Combining these energy equations with the swept-up mass term gives the generalized acceleration equation,
\begin{equation}
\frac{d\Gamma}{dt}
= \frac{\displaystyle \frac{dE}{dt} - \Gamma\mathcal{D}\frac{dE'_{\rm int}}{dt'} - (\Gamma^{2}-1)c^{2}\frac{dM_{\rm sw}}{dt}}
{\displaystyle M_{\rm ej}c^{2} + E'_{\rm int} + 2\Gamma M_{\rm sw}c^{2}},
\label{eq:dGdt_main}
\end{equation}
where $dM_{\rm sw}/dt = 4\pi R^{2} n_{\rm ext} m_{p}\, dR/dt$ is the swept-up mass rate for the isotropic ejecta (Appendix~\ref{app:shared_shock}).  Equations~(\ref{eq:dEdt_main})--(\ref{eq:dGdt_main}) show the direct causal path from engine parameters to the observable afterglow: changing $(B,P_0)$ changes $L_{\rm sd}(t)$, which changes $dE/dt$ and $dE'_{\rm int}/dt'$, and therefore changes the evolution of $R(t)$, $\Gamma(t)$, and $E'_{\rm int}(t)$.

Once the dynamical evolution $(\Gamma(t),R(t))$ is known, the post-shock magnetic field $B'$, the minimum electron Lorentz factor $\gamma_m$, and the characteristic synchrotron frequencies $(\nu_m,\nu_a)$ are computed from the standard external-shock relations (Appendix~\ref{app:ssa_dn}). The resulting synchrotron spectrum, including SSA and DN corrections, is then mapped to the observer frame to produce the multi-band light curves $F_\nu(t_{\rm obs})$ evaluated throughout this work.

While our code also computes the thermal ``merger-nova" (kilonova) component, the detection forecasts in this paper focus exclusively on the non-thermal afterglow channels.
\subsection{Isotropic Afterglow: Self-Absorption and Late-Time Corrections}
	\label{subsec:ejecta_afterglow}
	The ejecta drives a forward shock that accelerates electrons and strengthens magnetic fields, producing synchrotron radiation. Self-consistent treatment of SSA and DN effects is essential for providing robust sub-mm flux predictions. Sub-mm bands are particularly sensitive to the point where the emission shifts from being ``blocked" (optically thick) to ``transparent" (optically thin). We compute the observed flux $F_\nu(t_{\rm obs})$ by carefully tracking the self-absorption frequency $\nu_a$ and the characteristic frequency $\nu_m$. As the shock slows down, we apply DN corrections to the number of radiating electrons to ensure the late-time decay of the signal is physically accurate (Appendix \ref{app:ssa_dn}).  
	
	\subsection{Jet Afterglow and Viewing Angle}
	\label{subsec:jet_model}
	The jet component provides a bright early signal but is strongly dependent on the observer's viewing angle. When this module is active, we simulate a jet with a specific opening angle $\theta_j$ and integrate the emission across its surface. We adopt a two-sided jet, in which the true energy in one jet is \(E_{\rm jet}=E_{\rm iso}(1-\cos\theta_{j})/2\); the corresponding dynamics, optional lateral expansion, and the EATS arrival-time condition are documented in Appendix~\ref{app:jet}. Although the jet is much brighter than the isotropic ejecta at early times, it is only visible if the observer is looking nearly down the jet axis—a geometric constraint that we include in our overall rate estimates (Section~\ref{sec:rates}).

	\section{XSMT Sensitivity and Detectability Definition}
	\label{sec:xsmt_sens}
	
	To ground our physical predictions in observational reality, we must define a clear set of criteria for what constitutes a ``detectable" signal for the XSMT. In this section, we summarize the sensitivity model adopted for XSMT and the specific metrics used to evaluate the potential survey yield and to design an effective observing plan.
	
	\subsection{Adopted XSMT Thresholds}
	\label{subsec:xsmt_thresholds}
	
		Two angular scales are relevant for XSMT follow-up. The continuum camera has an instantaneous field of view (FoV) of $\sim 10'\times10'$ \citep{XSMT2025}, which sets the tiling efficiency for a GW localization region. By contrast, point-source association and source confusion are governed by the single-beam half-power beam width (HPBW). For a 15-m aperture, the diffraction estimate $\theta_{\rm HPBW}\simeq1.22\lambda/D$ gives $\theta_{\rm HPBW}\simeq22''$, $15''$, and $11''$ at 230, 345, and 460~GHz, respectively. Since representative LVK O4/O5 localization regions are expected to be of order tens of square degrees \citep{Abbott2020ObsScenarios}, far larger than the XSMT continuum FoV, XSMT is not a blind-search facility for large GW error regions; it requires prior localization to a host galaxy, optical/NIR counterpart, short-GRB afterglow, or a small galaxy-targeted search region before sub-mm follow-up can begin (see Section~\ref{sec:strategy}). Our analysis focuses on the three primary sub-mm continuum bands planned for XSMT, centered at 230, 345, and 460~GHz \citep{XSMT2025}. To make our forecasts as practical as possible, we adopt fixed sensitivity thresholds corresponding to a 5$\sigma$ detection in a 1-hour on-source integration:
	\begin{equation}
		F_{\rm th}(\nu)={1.5,\,2.9,\,10.2}\ {\rm mJy}\quad {\rm at}\quad \nu={230,345,460}\ {\rm GHz}.
		\label{eq:xsmt_thresholds}
	\end{equation}
	While the telescope's actual sensitivity can be improved by longer integration times (scaling roughly as $F_{\rm th} \propto t_{\rm int}^{-1/2}$), we use these 1-hour values as our standard baseline throughout the parameter sweeps and detectability maps. We adopt these fixed 5$\sigma$ (1 h) point-source thresholds as the baseline for our detectability criteria.

	\begin{table}[t]
		\centering
		\caption{XSMT $5\sigma$ (1 h) point-source thresholds used for detectability forecasts.}
		\label{tab:xsmt_thresholds}
		\begin{tabular}{ccc}
			\hline
			Band & $\nu$ (GHz) & $F_{\rm th}$ (mJy)\\
			\hline
			Band 6 & 230 & 1.5 \\
			Band 7 & 345 & 2.9 \\
			Band 8 & 460 & 10.2 \\
			\hline
		\end{tabular}
	\end{table}
	
	\subsection{Detectability Metrics and Interpretation}
	\label{subsec:detectability_metrics}
	For every model light curve $F_\nu(t_{\rm obs})$, we extract three core values: the peak flux density $F_{\rm pk}$, the time at which this peak occurs $t_{\rm pk}$ (observer-frame time measured from the onset of central-engine/ejecta evolution; this onset is taken to coincide with the merger/GW trigger), and the total duration the signal remains above the XSMT threshold, $t(F_\nu > F_{\rm th})$.  The primary criterion for detectability is the ``peak-flux condition", where a model is considered detectable if $F_{\rm pk} \ge F_{\rm th}$. For a model evaluated at a reference luminosity distance $D$, this peak-flux condition also defines the corresponding horizon distance,
	\begin{equation}
		D_{\max}(\nu)=D\sqrt{\frac{F_{\rm pk}(D,\nu)}{F_{\rm th}(\nu)}}.
		\label{eq:dmax}
	\end{equation}
	Here $F_{\rm pk}(D,\nu)$ is the model peak flux at observing frequency $\nu$, and $F_{\rm th}(\nu)$ is the adopted XSMT 5$\sigma$ threshold. This threshold serves as a conservative limit that directly determines the maximum distance (the horizon distance) at which a merger can be detected. Detectability maps are constructed by comparing model peak fluxes to these observational thresholds.  In our 2D maps (presented in Section~\ref{sec:survey}), the dashed contours indicate this detection boundary. Only parameter combinations within these contours represent events that XSMT could realistically characterize in standard 1-hour ToO visits. The ``time-above-threshold" metric is equally vital, as it helps us identify the optimal cadence for monitoring and alerts us to the temporal ``gap" that often exists between different emission components.  
	
	\subsection{Two-Channel Logic (Jet vs. Ejecta)}
	\label{subsec:two_channel_detectability}
	
	A key feature of our model is the potential for a ``bimodal" light curve when a relativistic jet is present. In such cases, a bright but very short peak (lasting only hours) is followed by a much broader peak from the isotropic ejecta (lasting months to years). By evaluating detectability separately for these two channels, we can tailor the XSMT strategy: the jet component requires a rapid-response ``Tier-1" visit, whereas the ejecta component allows for a more flexible, schedule-robust ``Tier-2" monitoring program. This distinction ensures that we maximize the science return for both beamed and isotropic signals (see Section~\ref{sec:strategy}).
	
	\section{Fiducial Predictions at $D=40$~Mpc}
	\label{sec:fiducial}
	To ground our analysis, we start by examining a baseline scenario for a local merger at $D=40$~Mpc, mirroring the distance of the landmark event GW170817 \citep{Hjorth2017}. For this fiducial case, we adopt parameters representative of a typical BNS merger: a magnetar engine with a dipole field $B=10^{15}$~G and an initial spin period $P_0=1$~ms, an ejecta mass $M_{\rm ej}=3\times10^{-3}M_\odot$, and a uniform external density $n_{\rm ext}=10^{-2}$~cm$^{-3}$. We evaluate the resulting light curves across the three XSMT bands and compare them to our 1-hour detection thresholds (Table \ref{tab:xsmt_thresholds}). The fiducial case anchors the interpretation of the subsequent broader parameter survey.
	
\subsection{Engine Dependence: Magnetar versus Black Hole}
\label{subsec:fid_engine_comparison}

\begin{figure*}[t]
    \centering
    \includegraphics[width=0.60\textwidth]{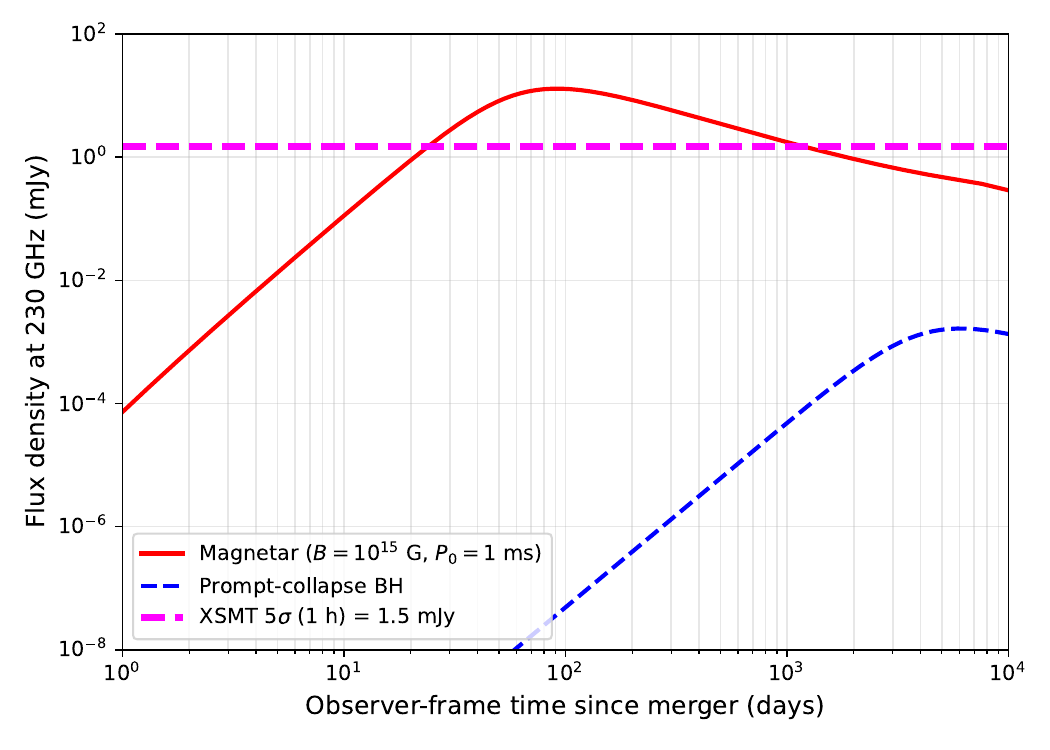}
	    \caption{Comparison of 230~GHz ejecta afterglow light curves for the fiducial magnetar engine (red solid) and a prompt-collapse BH (blue dashed). This contrast applies to the adopted fiducial ejecta, environment, and microphysical parameters. The horizontal dotted line marks the XSMT 1-hour $5\sigma$ threshold. The magnetar-boosted signal peaks at $\simeq13$~mJy and exceeds the threshold for $\sim24.2$--$1.18\times10^3$~days, while the BH-powered ejecta peaks at only $\simeq1.7\times10^{-3}$~mJy.}
    \label{fig:bh_vs_mag}
\end{figure*}

The central engine sets the energy budget before the observing strategy is considered. Figure~\ref{fig:bh_vs_mag} compares the 230~GHz ejecta afterglow for the fiducial magnetar engine with a prompt-collapse BH case ($L_{\rm eng}=0$), keeping the ejecta and environment fixed. This comparison is specific to the fiducial parameter set; the numerical contrast depends on the ejecta properties, ambient density, microphysical parameters, and any additional power supplied by fallback accretion onto the BH. The magnetar injects $\sim \xi E_{\rm rot} \approx 10^{52}$~erg over $T_{\rm sd}\approx 3.9\times 10^{3}$~s, re-energizing the ejecta and sustaining a higher shock velocity. The resulting signal peaks at $F_{\rm pk}\approx 13$~mJy, exceeds the XSMT 1-hour $5\sigma$ threshold for $\sim 24.2$--$1.18\times10^3$~days, and reaches $D_{\max}\approx 118$~Mpc.

By contrast, the BH-powered ejecta evolves without sustained injection. Its afterglow peaks at only $F_{\rm pk}\approx1.7\times10^{-3}$~mJy, a factor of $\sim 8\times10^{3}$ below the magnetar case, and remains undetectable at 40~Mpc; the corresponding horizon distance is only $D_{\max}\approx1.3$~Mpc. This comparison establishes the main physical point of the fiducial calculation: a multi-epoch XSMT detection of an ejecta afterglow at $\sim 100$~Mpc, with a week-to-month rise and an SSA-consistent spectrum, would favor long-lived engine energy injection. A non-detection alone, however, would not uniquely imply prompt collapse because distance, low density, cadence limitations, weather, or less favorable microphysical parameters could also suppress the signal.

	\subsection{Persistence of the Isotropic Ejecta Afterglow}
	\label{subsec:fid_ejecta_only}
	Figure \ref{fig:fiducial_lc} (left panel) presents the evolution of the isotropic ejecta afterglow. At 230~GHz, the signal rises steadily over weeks to months, peaking at $t_{\rm pk}\simeq 92$~days with a flux density that remains well above the XSMT threshold for an extended period. This long-term persistence, spanning months to years, implies that the ejecta channel is notably robust to scheduling constraints. Even with significant delays following a GW trigger, XSMT remains likely to capture the signal during its broad peak or slow decay. The isotropic ejecta afterglow offers a long and schedule-robust window for sub-mm follow-up. While the behavior at 345~GHz is qualitatively similar, it features a tighter observation window; Band 8 (460~GHz) is generally less favorable for the ejecta component at this distance unless the integration time is significantly increased.
	
	\subsection{Bimodal Evolution and the Jet-Ejecta Gap}
	\label{subsec:fid_jet_plus_ejecta}
\begin{figure*}[t]
		\centering
		\includegraphics[width=0.45\textwidth]{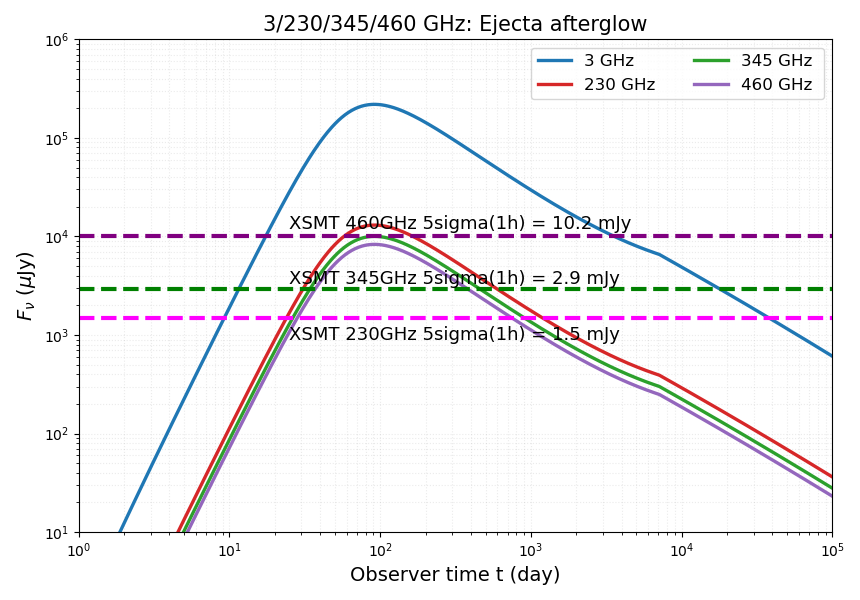}
		\includegraphics[width=0.45\textwidth]{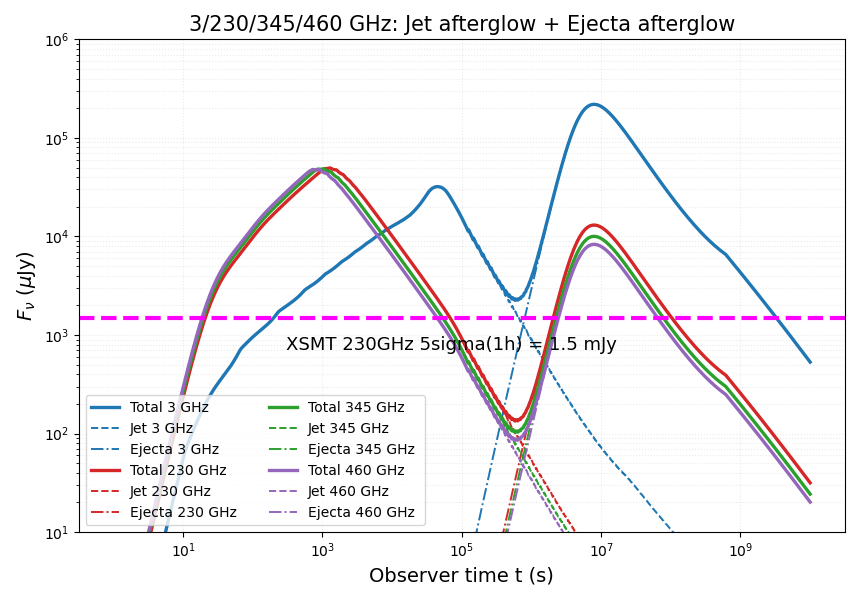}
		\caption{Fiducial sub-mm light curves at $D=40$~Mpc for 230/345/460~GHz.
			Solid curves show total flux. Dashed horizontal lines mark the adopted XSMT 5$\sigma$(1~h) thresholds (Table~\ref{tab:xsmt_thresholds}).
			Left: ejecta-only case (isotropic external-shock afterglow). Right: jet+ejecta case (top-hat jet, on-axis).
			The jet component produces an early bright peak (hours) with a short detectability window, whereas the ejecta component provides a late-time peak (months) with a long detectability window at 230~GHz.}
		\label{fig:fiducial_lc}
	\end{figure*}

	The inclusion of a relativistic jet component introduces a dramatic, early-time signature. As shown in Figure \ref{fig:fiducial_lc} (right panel), an on-axis top-hat jet generates an intense sub-mm peak within hours ($t \sim 10^{-2}$~days) of the merger, followed by a steep decline. This creates a characteristic ``bimodal" light curve, where the early jet-driven peak and the late ejecta-driven peak are separated by a distinct temporal gap where the flux may drop below detection limits.  This clear separation in time---hours versus months---dictates our follow-up logic. Detecting the jet requires an immediate response, while the ejecta afterglow allows for a more measured monitoring program. Jet detectability demands rapid ToO response, whereas the later ejecta peak is more tolerant of scheduling delays. We note that while the jet is intrinsically much brighter, its contribution to the overall detection rate is limited by its directional nature, whereas the ejecta is visible from any angle.  
	
\subsection{Fiducial Detectability Summary}
	\label{subsec:fid_summary}
	
	Table \ref{tab:jet_ejecta_main} summarizes the peak fluxes, detection windows, and maximum horizon distances for both channels. The 230~GHz band emerges as our primary workhorse, offering the largest detection volume and the longest window for the isotropic channel. The 345~GHz band provides crucial spectral information for nearby events or as a follow-up to a confirmed 230~GHz detection. In contrast, the 460~GHz band is most effectively reserved for very nearby mergers or rapid ``jet chasing" under optimal weather conditions. Despite the brightness of the jet, the isotropic channel remains the more reliable detection route for XSMT.

	\section{Parameter-space Survey and Detectability Maps}
	\label{sec:survey}

	Beyond the fiducial scenario, the potential for XSMT to detect a sub-mm counterpart depends on the vast diversity of merger conditions. Here we quantify how the light curves respond to changes in the central engine and the surrounding environment. We explore these dependencies through: (i) one-dimensional (1D) sweeps that isolate the impact of individual parameters and (ii) two-dimensional (2D) detectability maps that identify the specific regions of parameter space where XSMT can realistically expect a detection. We explore physically motivated ranges spanning multiple orders of magnitude in engine and environmental parameters.

\subsection{Scan Design and Baseline Assumptions}
\label{subsec:scan_design}
Unless otherwise noted, we vary one parameter at a time while holding others at their fiducial values (Section~\ref{sec:fiducial}). For magnetar-driven models, we focus on the $(P_0, B)$ plane, as these variables govern the total energy reservoir and the duration of injection. For the external-shock evolution, we examine the $(M_{\rm ej}, n_{\rm ext})$ plane, which dictates the deceleration timescale and the density of radiating electrons. All resulting light curves are evaluated at 230, 345, and 460 GHz against the fixed 1-hour $5\sigma$ thresholds (Table \ref{tab:xsmt_thresholds}), represented as dashed reference lines in our plots.

\subsection{One-dimensional Sweeps: Key Physical Trends}
\label{subsec:one_d_sweeps}
\begin{figure*}[t]
			\centering
			\begin{subfigure}[b]{0.7\textwidth}
				\centering
				\includegraphics[width=\linewidth]{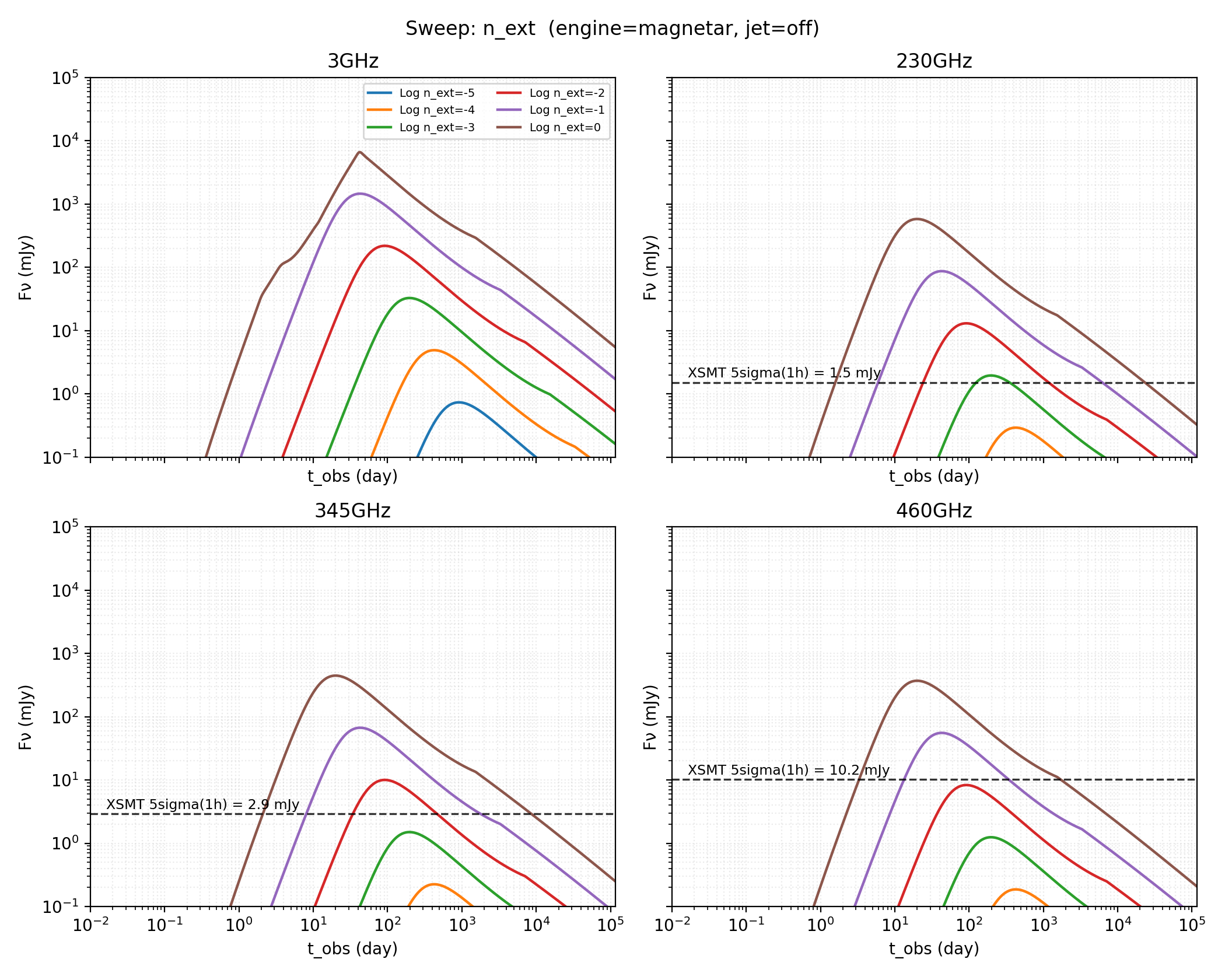}
				\caption{1D sweep for ambient medium density $n_{\rm ext}$.}
				\label{fig:sweeps:a}
			\end{subfigure}
			\vspace{1em}
			\begin{subfigure}[b]{0.7\textwidth}
				\centering
				\includegraphics[width=\linewidth]{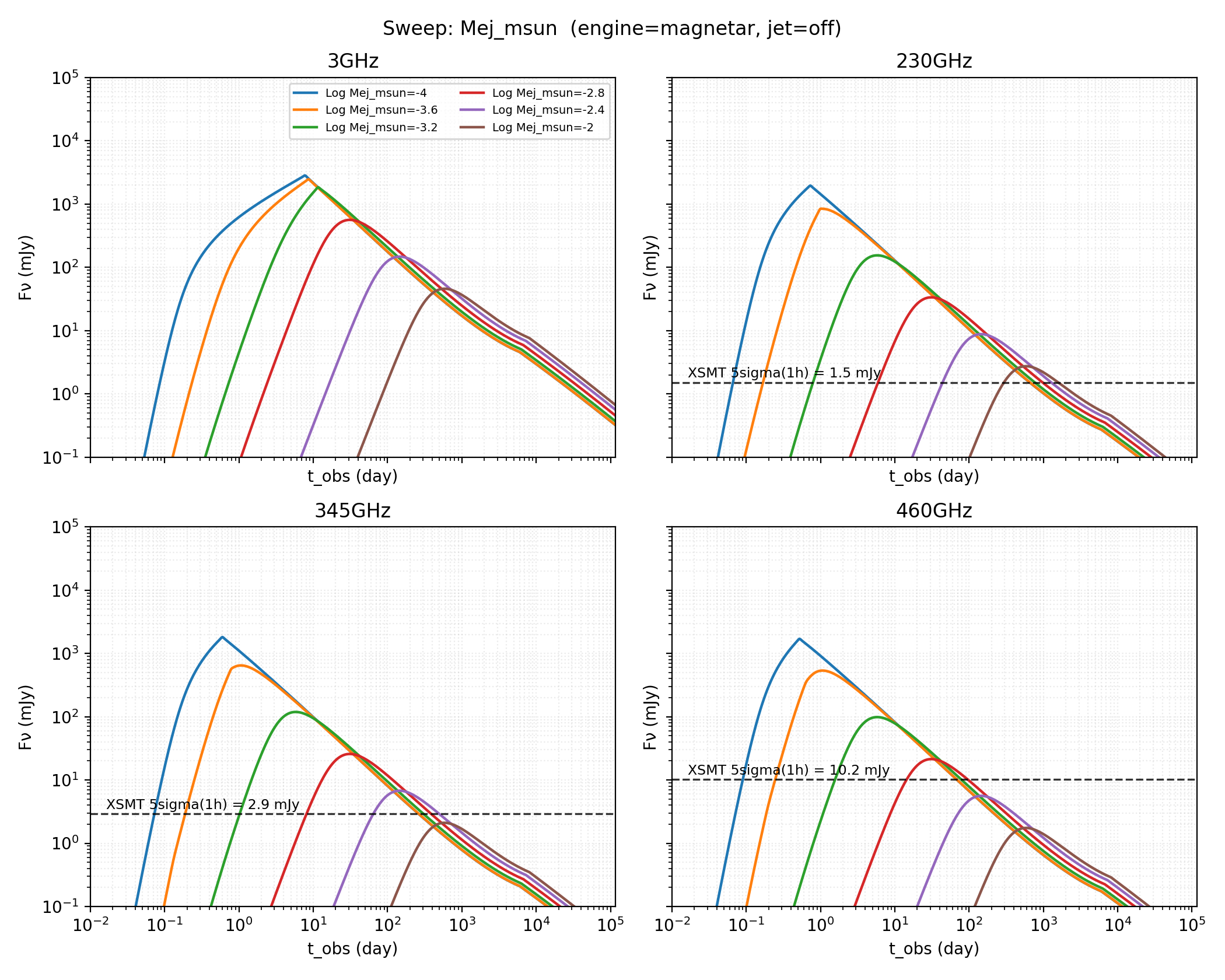}
				\caption{1D sweep for ejecta mass $M_{\rm ej}$}
				\label{fig:sweeps:b}
			\end{subfigure}
			\caption{Representative 1D sweeps of sub-mm light curves at 230/345/460~GHz (and 3~GHz for context), assuming a magnetar central engine and the jet contribution turned off. Dashed horizontal lines denote the fixed XSMT 5$\sigma$(1~h) thresholds. Each panel shows how peak flux and detectability window respond to varying a single physical parameter while holding others at fiducial values.}
			\label{fig:sweeps}
		\end{figure*}

Figure \ref{fig:sweeps} illustrates representative sweeps, from which several robust trends emerge regarding the ``detectability window" of the signal.

	\paragraph{The Role of Ambient Density.}
	The environment density, $n_{\rm ext}$, acts as the primary gatekeeper for sub-mm detectability. A higher density increases the number of available electrons and strengthens the magnetic field behind the shock, leading to significantly higher peak fluxes and earlier peak times. Because the flux response is so steep, the uncertainty in the local circumburst environment remains the largest contributor to the systematic uncertainty in our rate forecasts (Section~\ref{sec:rates}). Sub-mm detectability is most sensitive to the ambient density and the central engine's total energetics.
	
	\paragraph{Ejecta Mass and Deceleration.}
	Increasing the ejecta mass $M_{\rm ej}$ generally delays the point at which the outflow begins to slow down, shifting the afterglow peak to later times. While a larger mass can provide more material for radiation, its primary effect in the sub-mm band is often seen in the timing and duration of the visibility window rather than a simple linear evolution in brightness.  
	
\subsection{Two-dimensional Detectability Maps}
	\label{subsec:two_d_maps}
	
	To provide a comprehensive forecast, we construct 2D maps that visualize the peak flux ($F_{\rm pk}$) and peak time ($t_{\rm pk}$) across different regimes. Here we present the ejecta-only case for illustrative purposes. The dashed contour marks the detectability boundary defined by $F_{\rm pk}=F_{\rm th}$ (Section~\ref{subsec:detectability_metrics}). 
	
\begin{figure*}[t]
		\centering
		\includegraphics[width=0.48\textwidth]{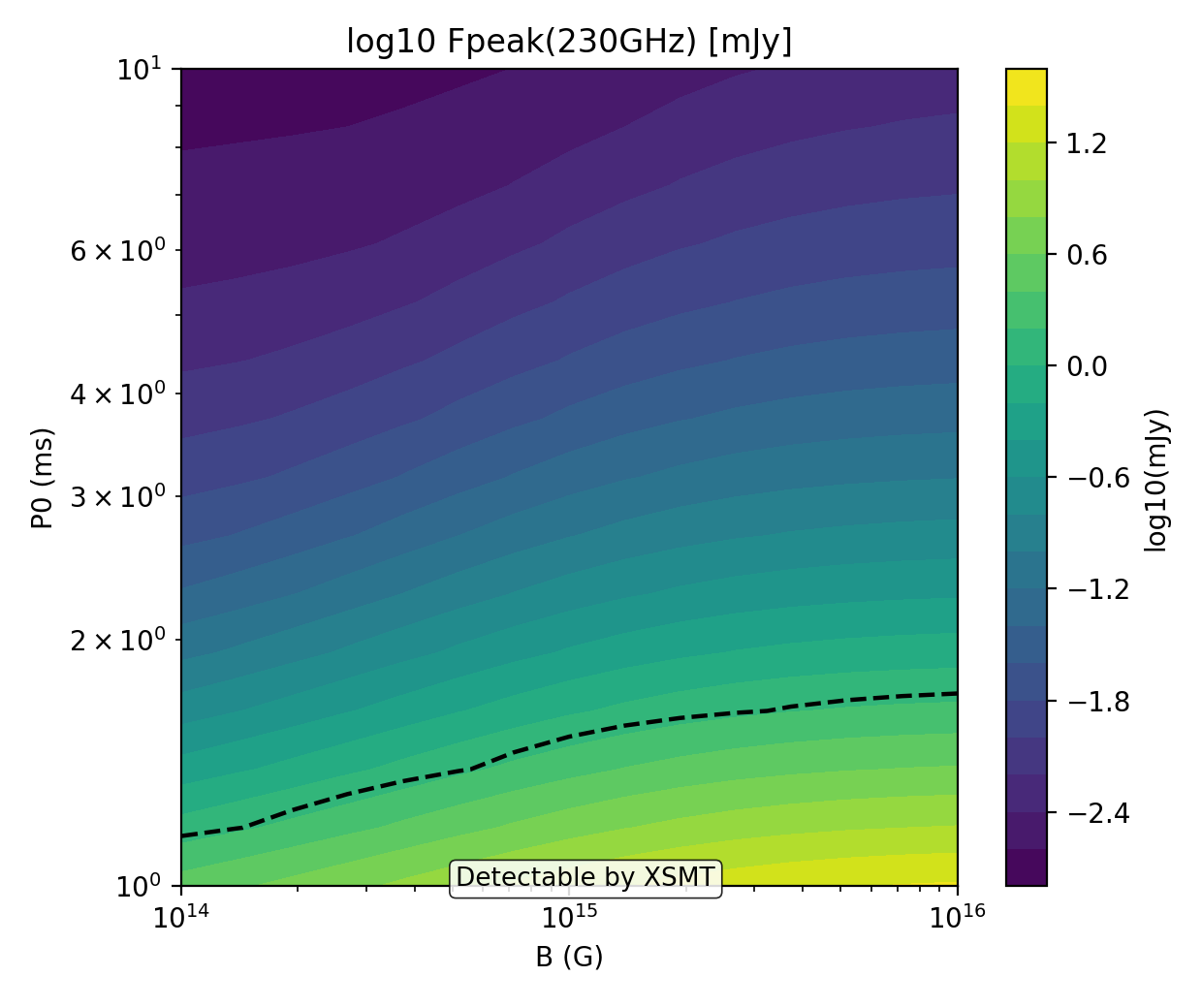}
		\includegraphics[width=0.48\textwidth]{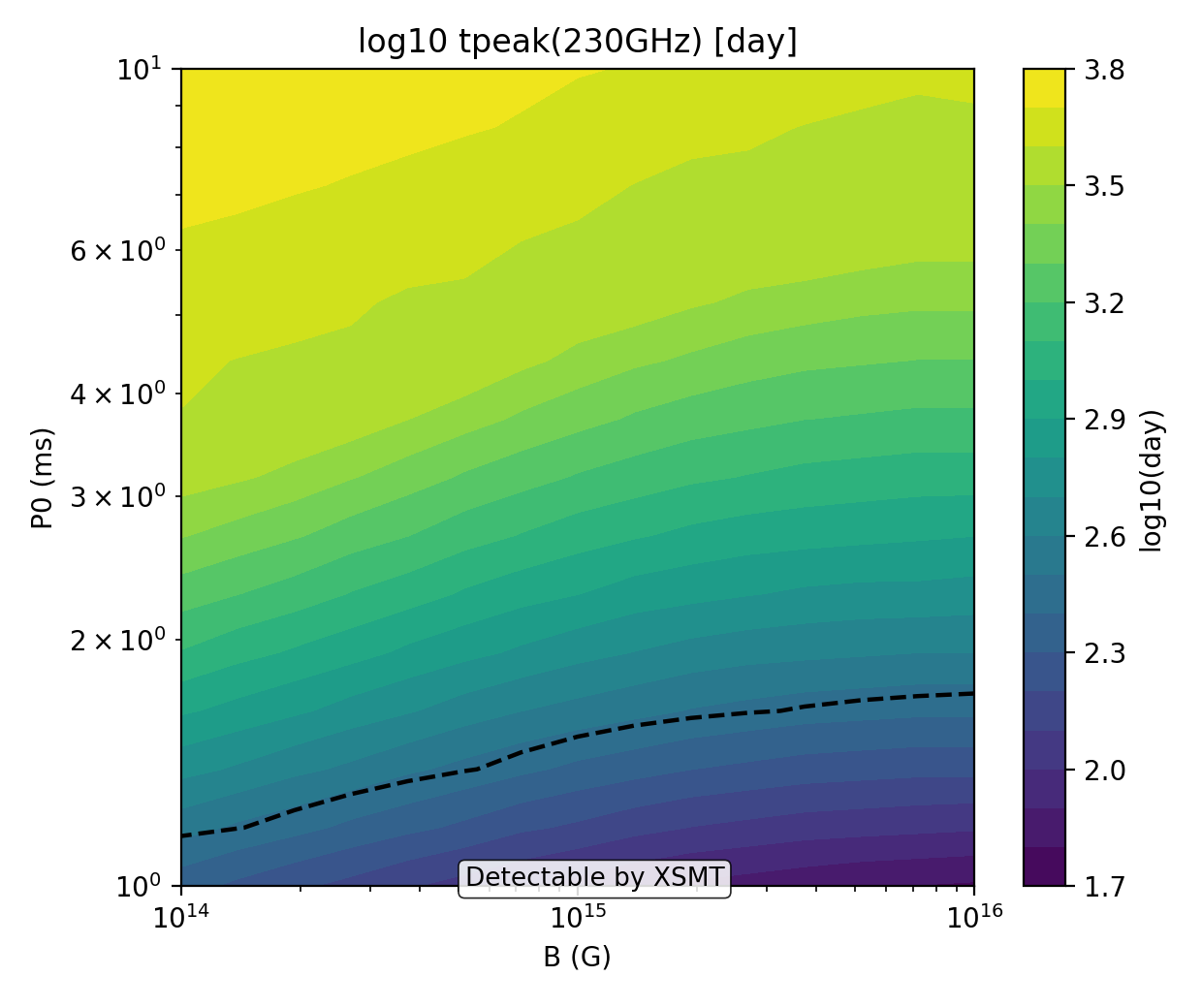}
		\caption{Detectability maps in the magnetar-engine plane $(P_0,B)$ at 230~GHz. Colors show $\log_{10}F_{\rm pk}$ (mJy) and $\log_{10}t_{\rm pk}$ (day). The dashed contour marks the boundary $F_{\rm pk}=F_{\rm th}$, where $F_{\rm th}$ is the fixed XSMT 5$\sigma$(1~h) threshold. Only within the dashed contour is the peak time observationally relevant for 1~h integrations.}
		\label{fig:maps_P0B}
	\end{figure*}

	\paragraph{The Magnetar Plane ($P_0, B$).}
	For magnetar engines, shorter rotation periods ($P_0$) and stronger magnetic fields ($B$) provide a more powerful energy boost to the ejecta. This injection not only brightens the signal but can also alter the timing of the peak depending on how long the engine remains active. Figure \ref{fig:maps_P0B} shows that the region detectable by XSMT expands significantly toward the lower-right of the $(P_0, B)$ plane—representing rapid rotators with strong dipole fields. Interestingly, even within the detectable region, the peak times can vary from weeks to over a year, suggesting that a successful monitoring program must be prepared for a wide range of evolution speeds. XSMT's detectability occupies a well-defined region in the $P_0-B$ and $M_{ej}-n_{\rm ext}$ parameter planes.
	
\begin{figure*}[t]
		\centering
		\includegraphics[width=0.48\textwidth]{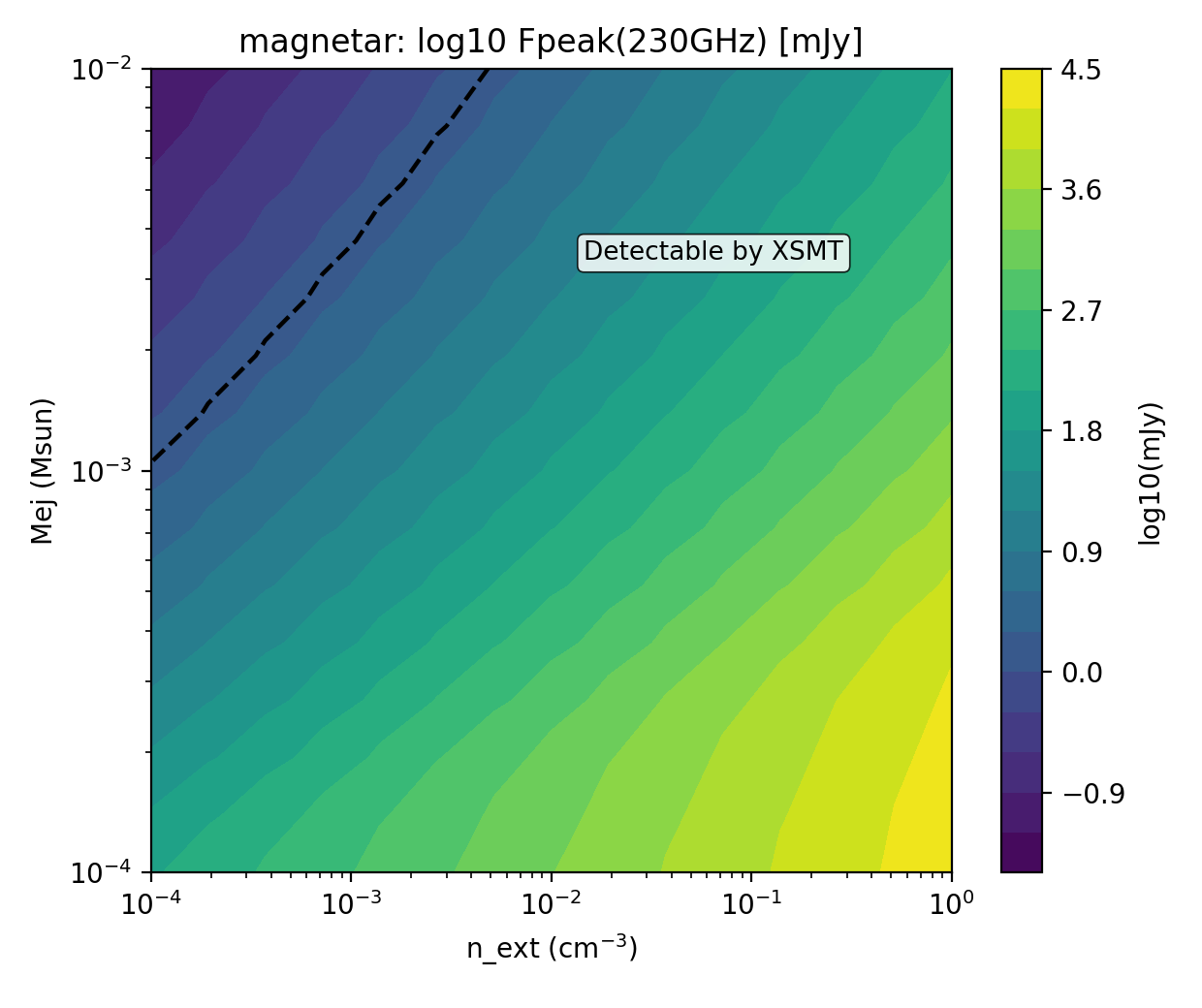}
		\includegraphics[width=0.48\textwidth]{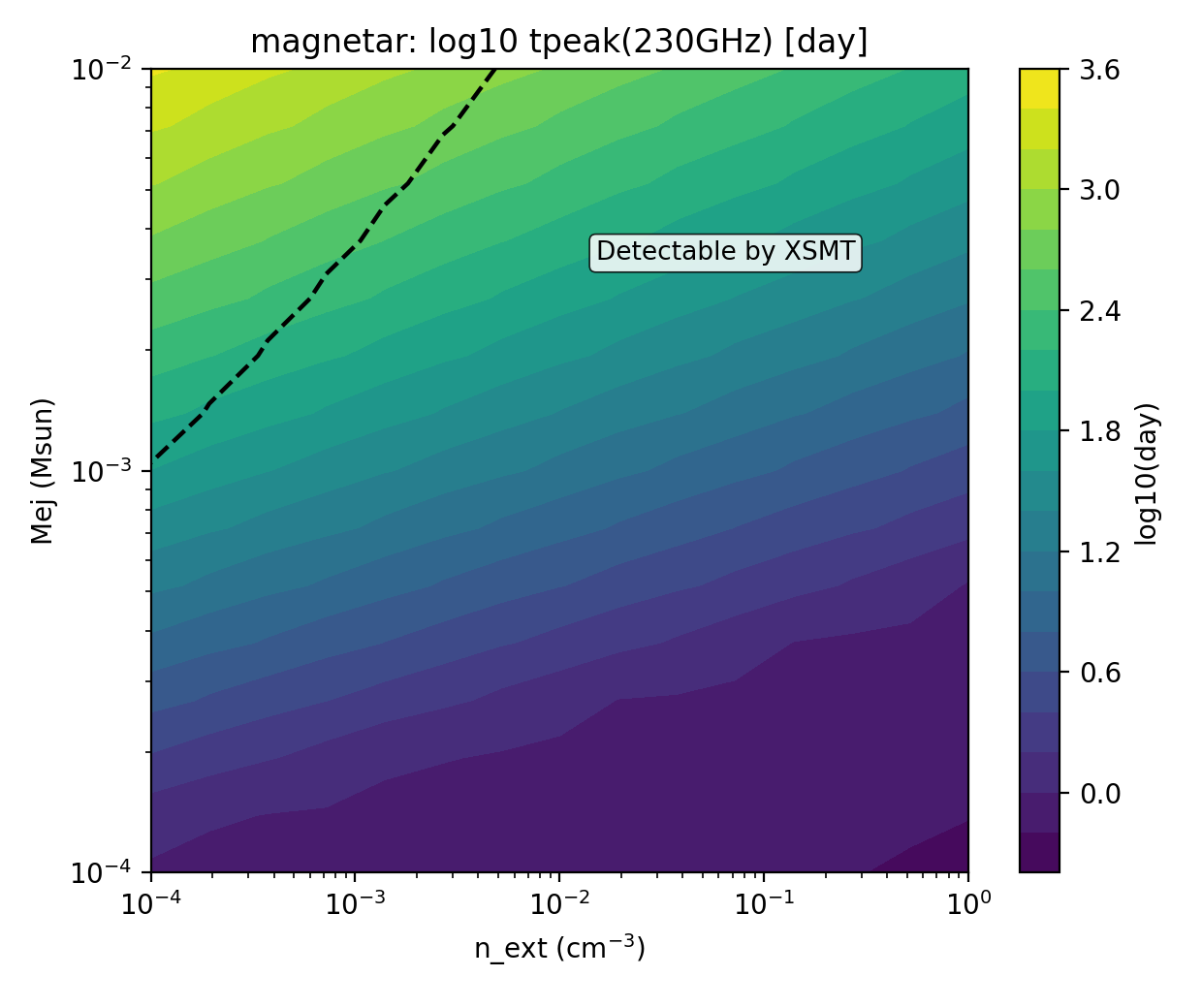}
		\caption{Detectability maps in the external-shock plane $(M_{\rm ej},n_{\rm ext})$ at 230~GHz. The dashed contour denotes $F_{\rm pk}=F_{\rm th}$. The maps highlight the strong sensitivity of sub-mm detectability to $n_{\rm ext}$ and the role of $M_{\rm ej}$ in setting the peak time and thus the optimal follow-up window.}
		\label{fig:maps_Mejn}
	\end{figure*}

	\paragraph{The External-shock plane ($M_{\rm ej}, n_{\rm ext}$).}
	In Figure \ref{fig:maps_Mejn}, the detectability boundary is clearly dominated by $n_{\rm ext}$, while $M_{\rm ej}$ primarily controls when that peak will occur. This map serves as a direct bridge to our observing strategy: for events in low-density environments, XSMT should prioritize late-time monitoring, whereas high-density cases require earlier observations to capture the brighter, faster-evolving peaks. 
	
	\subsection{Prioritizing Bands Based on Maps}
	\label{subsec:band_priorities}
	
		Across all surveyed parameters, the 230 GHz band consistently offers the largest detectable area, making it the main discovery band. The 345 GHz band is valuable for events that fall comfortably within the detection contour, providing spectral leverage: when paired with 230~GHz, the 230--345~GHz spectral slope tests whether the sub-mm bands are already in the optically thin synchrotron regime or show curvature associated with the SSA turnover, helping to break degeneracies between shock energy, ambient density, and microphysics. These detectability maps translate directly into a frequency-prioritized follow-up plan for future GW triggers.

	\section{Detection Rates}
	\label{sec:rates}
	
		The ultimate value of sub-mm follow-up lies in the number of events a facility can realistically characterize. Here we translate our detectability results into expected detection yields for XSMT, treating the isotropic ejecta and the beamed jet as two separate discovery channels. Throughout this calculation, we adopt a BNS volumetric merger rate of $\mathcal{R}_{\rm BNS}=7.6$--250~Gpc$^{-3}$~yr$^{-1}$ \citep{Abac2025}. It is important to note that these estimates scale linearly with the merger rate and would be further modulated by the fraction of mergers that produce a long-lived engine.  
	
	\subsection{Horizon Distance and Effective Volume}
	\label{subsec:dmax_vmax}
	
	To estimate the event rate, we use the horizon distance $D_{\max}$ defined by the peak-flux criterion in Equation~(\ref{eq:dmax}). This horizon distance yields an effective comoving volume, $V_{\max}(\nu) = \frac{4\pi}{3} D_{\max}^3$. The rate estimate follows directly from the horizon volume scaled by the BNS merger rate density.
	
	\subsection{Yield from the Isotropic Ejecta Channel}
	\label{subsec:rate_ejecta}
	
		The isotropic ejecta afterglow is our most reliable detection channel because its visibility does not depend on the merger's orientation. The all-sky detection rate is simply $\dot N_{\rm ej}(\nu)=\mathcal{R}_{\rm BNS}\,V_{\max}^{\rm ej}(\nu)$. In this channel, the primary uncertainties are astrophysical rather than geometric. Throughout our rate calculation we assume a magnetar engine fraction $f_{\rm mag}=1$ (i.e., every BNS merger produces a long-lived magnetar); the quoted rates therefore represent upper limits, and the true rate scales linearly with $f_{\rm mag}$. As our earlier sweeps demonstrated, the ambient density $n_{\rm ext}$ is the dominant factor controlling the peak brightness. Environmental uncertainty is the leading systematic in the rate.  For our fiducial magnetar case at 230 GHz, the horizon reaches $D_{\max}\approx118$~Mpc, leading to an expected detection rate of $0.052$--1.72~yr$^{-1}$ under current BNS merger-rate priors. While the 345 GHz and 460 GHz bands offer smaller horizons, they remain essential for confirming the spectral shape of the nearest events. Overall, 230 GHz dominates the expected yield for XSMT.
	
	\subsection{Yield from the Beamed Jet Channel}
	\label{subsec:rate_jet}
	
	The jet afterglow can be orders of magnitude brighter than the ejecta at early times, potentially pushing the horizon distance far beyond 200 Mpc. However, this brightness is counteracted by the narrowness of the beam. We estimate a conservative on-axis rate by including a geometric beaming factor, $f_b \simeq 1-\cos\theta_j \approx \theta_j^2/2$:
	\begin{equation}
		\dot N_{\rm jet,on}(\nu)\approx \mathcal{R}_{\rm BNS}\,V_{\max}^{\rm jet}(\nu)\,f_b.
		\label{eq:rate_jet_final}
	\end{equation}
	For a typical opening angle of $\theta_j=0.1$ rad, the beaming factor is small ($\sim 0.005$), meaning that only a tiny fraction of mergers will have their jets pointed directly at Earth. Consequently, the jet rate is suppressed by geometry despite its much higher intrinsic brightness. We consider this estimate conservative, as real jets may have a ``structured" edge that allows for detection at slightly off-axis angles, and GW selection effects naturally favor events with lower inclination.
	
	\subsection{Summary of Detection Potential}
	\label{subsec:rate_summary}
	
	The comparison between these two channels, summarized in Table \ref{tab:jet_ejecta_main}, clarifies the operational priorities for XSMT. The isotropic ejecta channel is the primary route for building a statistical sample of sub-mm counterparts due to its larger effective yield and long visibility window. The jet channel, while rarer, provides a high-signal-to-noise opportunity for rapid-response science on a subset of high-priority triggers. This dual-track nature of the sub-mm signal is what makes the two-tier ToO strategy, discussed in the next section, so essential.
	
	\begin{deluxetable*}{cccccccc}
			\caption{Fiducial detectability and rates at $D=40$~Mpc. Fiducial parameters: $B=10^{15}$~G, $P_0=1$~ms, $M_{\rm ej}=3\times10^{-3}M_\odot$, $n_{\rm ext}=10^{-2}$~cm$^{-3}$. Rates are scaled to $\mathcal{R}_{\rm BNS}=7.6$--250~Gpc$^{-3}$~yr$^{-1}$. Ejecta-channel rates assume a long-lived magnetar fraction $f_{\rm mag}=1$ and scale linearly with this fraction. Jet rates include beaming factor $f_b \approx 0.005$ for $\theta_j=0.1$~rad. \label{tab:jet_ejecta_main}}
		\tablecolumns{8}
		\tablewidth{0pt} 
		\tablehead{
			\colhead{Band} & \colhead{$F_{\rm th}$} & \colhead{Comp.} & \colhead{$F_{\rm pk}$} & 
			\colhead{$t_{\rm pk}$} & \colhead{Window} & \colhead{$D_{\max}$} & \colhead{$\dot{N}$} \\
			\colhead{(GHz)} & \colhead{(mJy)} & \colhead{} & \colhead{(mJy)} & 
			\colhead{(day)} & \colhead{(day)} & \colhead{(Mpc)} & \colhead{(yr$^{-1}$)}
		}
		\startdata
		230 & 1.5  & Jet & 49.6 & $1.47\times10^{-2}$ & $2.5\times10^{-4}$--0.77 & 230 & $1.9\times10^{-3}$--$0.064$ \\
			&      & Ej. & 13.0 & 91.6               & 24.2--$1.18\times10^3$               & 118 & $0.052$--$1.72$ \\
		\hline
		345 & 2.9  & Jet & 48.6 & $1.18\times10^{-2}$ & $3.2\times10^{-4}$--0.31 & 164 & $7.0\times10^{-4}$--$0.023$ \\
		&      & Ej. & 10.0 & 91.6               & 34.2--459               & 74.4 & $0.013$--$0.43$ \\
		\hline
		460 & 10.2 & Jet & 47.9 & $8.48\times10^{-3}$ & $9.5\times10^{-4}$--0.07 & 86.7 & $1.0\times10^{-4}$--$3.4\times10^{-3}$ \\
		&      & Ej. & 8.31 & 91.6               & ---                   & 36.1 & $1.5\times10^{-3}$--$0.049$ \\
		\enddata
	
	\end{deluxetable*}

	\section{Observing Strategy for XSMT}
	\label{sec:strategy}
	
	To maximize the scientific impact of XSMT, our model predictions must be translated into an observing program that is both executable and diagnostically powerful. The fundamental challenge lies in the temporal bimodality of the signal: the relativistic jet offers a bright but short-lived window of only hours, while the isotropic ejecta provides a steady signal spanning weeks to years that dominates the overall detection rate. A two-tier ToO program maximizes the science return per observing hour by targeting these distinct components.  
	
	\subsection{Tier-1: Rapid Jet-Chasing ToO (Hours)}
	\label{subsec:tier1}
	
	The jet component can exceed XSMT thresholds by orders of magnitude in favorable geometries, but its visibility is fleeting. In our fiducial case, the above-threshold window lasts $\lesssim 1$ day at 230~GHz and only a few hours in the higher frequency bands (Table~\ref{tab:jet_ejecta_main}). Tier-1 is therefore designed as a high-speed response to capture this early-time signature.  
	
	\paragraph{Prioritization and Trigger Selection.}
		Because the continuum FoV and HPBW defined in Section~\ref{subsec:xsmt_thresholds} imply limited areal survey speed but useful beam-scale association, XSMT should be treated as a targeted follow-up facility rather than a blind-search instrument for large GW error regions. We recommend activating Tier-1 only when a secure EM localization is available—such as a detected short GRB, a high-energy afterglow, or a small-enough GW localization volume to allow for a targeted galaxy-catalog search. High-priority candidates include events where the GW inclination posterior favors a near-axis geometry, significantly increasing the probability of a bright jet detection.  
	
	\paragraph{Recommended Tier-1 Cadence.}
	A minimal rapid sequence targets the jet's rise and early decay:
	\begin{itemize}
		\item $T_0 + \lesssim 6$ h: Simultaneous or rapid-offset observations at 230~GHz and 345~GHz (1~h each);
		\item $T_0 + \sim 12$ h: Repeat 230~GHz (1~h) to establish the initial decay slope; if still within the expected jet window and conditions permit, 345~GHz (1~h);
		\item High-Priority Add-on: If conditions are excellent, Band 8 (460~GHz) observations within the first few hours can provide critical data on the high-frequency spectral slope before the jet fades.
	\end{itemize}  
	
	\subsection{Tier-2: Long-Baseline Ejecta Monitoring (Weeks to Years)}
	\label{subsec:tier2}
	
	Tier-2 targets the isotropic ejecta afterglow, which represents the most reliable discovery channel for XSMT. Because this component is visible from all angles, it allows for a more flexible scheduling cadence. At 230~GHz, the signal typically remains detectable for months to years, allowing delayed or repeated observations to characterize the merger's total energy budget and environment.  
	
	\paragraph{Monitoring Cadence and Multi-band Synergy.}
	We propose a sparse, logarithmic cadence centered on the expected $t_{\rm pk}\simeq 92$~d peak:
	\begin{itemize}
		\item Rise and Peak (30, 60, and 90--120 d): 230~GHz (1~h) to track the shock's growth.
			\item Peak and Decay (150 and 250 d): 230~GHz (1~h) to constrain the late-time decay and DN evolution.
	\end{itemize}
	We recommend adding 345~GHz observations at the 60-day and 120-day marks for events within $D \lesssim 75$~Mpc or once a 230~GHz detection is secured. Combining these frequencies provides spectral leverage on the sub-mm slope and can reveal curvature toward the SSA turnover when it lies near the observing bands, helping to reduce degeneracies between ambient density and microphysics.  
	
	\subsection{Operational Logic: Navigating the Jet-Ejecta Gap}
	\label{subsec:gap_logic}
	
	One of the most practical insights from our modeling is the existence of a temporal ``gap" between the jet and ejecta signals. If Tier-1 observations result in a non-detection, XSMT resources can be preserved by pausing follow-up until the predicted onset of the ejecta window ($\sim 25$--35 days for the fiducial 230--345~GHz bands). This gap-aware cadence avoids repeated observations during the interval when neither component is expected to be bright, while still sampling the later ejecta rise.

	\section{Discussion}
	\label{sec:discussion}
	
		Our forecasts and strategy recommendations are designed to be facility-specific and actionable for the XSMT era. Here we place the calculation in the broader context of previous theoretical work and existing radio/sub-mm facilities, then discuss the leading sources of uncertainty, the impact of jet geometry and selection effects, frequency prioritization, multiwavelength synergies, source confusion, and operational limitations.

\subsection{Comparison with Previous Theoretical Studies}
\label{subsec:comparison_theory}

Previous studies have shown that long-lived merger remnants can enhance late-time radio/sub-mm emission, but they usually target different diagnostics. PWN-focused calculations \citep[e.g.,][]{Ren2022,Mukhopadhyay2025,Omand2025} use the internal wind-powered nebula or ejecta-PWN system as the remnant signature. Our calculation instead uses the external forward shock of the bulk ejecta as the primary XSMT observable, with the magnetar entering through energy injection into the ejecta dynamics via $L_{\rm eng}(t)$ rather than through a separately modeled PWN radiative component.

\citet{Sarin2022} provide the closest ejecta-focused comparison. They studied how magnetar spin-down, gamma-ray leakage, and the partition of engine power between thermal emission and kinetic acceleration shape the diversity of magnetar-powered kilonovae and their radio/X-ray kilonova afterglows. Our work shares the same basic premise that sustained engine injection can accelerate the ejecta and brighten the later external-shock afterglow, but it differs in how the late-time dynamics are evolved.

In \citet{Sarin2022}, their Eq.~(6) follows the engine-accelerated kilonova ejecta using the fixed ejecta mass and internal-energy budget. This captures the early engine-acceleration phase, but it does not explicitly include the swept-up circumburst mass $M_{\rm sw}$ or the associated deceleration term $dM_{\rm sw}/dt$. By contrast, our acceleration equation (Eq.~(\ref{eq:dGdt_main}); Appendix~\ref{app:ejecta_dyn}) includes both the engine/radioactive/radiative energy-balance terms in the numerator and the inertia of the original ejecta, internal energy, and swept-up external material in the denominator. The same calculation therefore follows the outflow from early engine acceleration to late external-shock deceleration and, eventually, the DN regime used in the synchrotron calculation.

The novelty of the present work is therefore not the general idea that magnetar remnants can brighten merger afterglows. It is the facility-specific translation of this early-to-late dynamical framework into XSMT observables: peak fluxes, peak times, time-above-threshold windows, horizon distances, rates, and a concrete monitoring cadence at 230/345/460~GHz. In short, previous studies establish the physical plausibility of magnetar-enhanced afterglows; here we ask when XSMT can detect and monitor the external-shock signal well enough to distinguish sustained engine injection from a prompt-collapse BH scenario.

\subsection{Complementarity with Existing Facilities}
\label{subsec:facility_comparison}

To place XSMT in the broader landscape of sub-mm and radio facilities, we compare its continuum sensitivity with ALMA and the VLA for a fiducial 1-hour integration at $5\sigma$. Representative ALMA continuum values from the official ALMA Sensitivity Calculator and Technical Handbook give $5\sigma\approx 0.15$~mJy at 230~GHz (Band~6) and $5\sigma\approx 0.25$~mJy at 345~GHz (Band~7), depending on weather and setup \citep{ALMASensitivityCalculator,ALMATechnicalHandbook}. These are roughly an order of magnitude deeper than the XSMT 1-hour thresholds of 1.5 and 2.9~mJy in the same bands. For the VLA, the NRAO Observational Status Summary gives the thermal-noise formalism and representative SEFD values; adopting an 8~GHz continuum bandwidth and 1~h on source gives approximate $5\sigma$ sensitivities of $\sim0.02$~mJy in Ka band (33~GHz) and $\sim0.04$~mJy in Q band (45~GHz) \citep{VLAOSS}. The VLA therefore provides much deeper low-frequency coverage, but it does not sample the 230--460~GHz sub-mm spectral regime probed by XSMT and ALMA.

The facility comparison suggests a practical division of labor. ALMA is the deepest instrument for detailed sub-mm spectral characterization of nearby events, and the VLA is important for anchoring the lower-frequency synchrotron slope. These facilities, however, are heavily subscribed general-purpose observatories, so repeated long-baseline monitoring of every GW trigger may be difficult in practice. XSMT is less sensitive per pointing, but it can provide sustained 230--460~GHz monitoring and catch the broad ejecta afterglow peak. ALMA and the VLA can then supply deeper observations at selected epochs, especially near the predicted peak or after an XSMT detection, to refine the SED and constrain the optically thin slope, possible SSA curvature, and microphysical parameters. Joint XSMT+ALMA+VLA campaigns would therefore combine cadence, depth, and frequency coverage in a way that is well suited to testing whether a detected ejecta afterglow requires long-lived engine injection.

\subsection{Dominant Uncertainties and Parameter Degeneracies}
	\label{subsec:uncertainties}
	
	The leading astrophysical uncertainty for sub-mm detectability is the circumburst density environment $n_{\rm ext}$. As demonstrated by our surveys, $n_{\rm ext}$ acts as a ``gatekeeper" that modulates the post-shock energy density and the magnetic field strength, thereby shifting both the peak flux and the SSA frequency. A shift in $n_{\rm ext}$ by one order of magnitude can alter the inferred horizon distance and the resulting event rate by a factor of several, making environmental priors the primary systematic in any sub-mm forecast.  Furthermore, microphysical parameters $(\epsilon_e, \epsilon_B, p)$ introduce degeneracies that can mimic the signatures of different energy-injection histories. For instance, a high-density environment with low magnetic-field efficiency can produce light curves similar to those of a low-density environment with a powerful magnetar engine. Multi-band sub-mm measurements are therefore needed to test the optically thin slope and search for spectral curvature, rather than relying on a single-band flux normalization. Environmental uncertainty is the leading systematic in the rate estimates, and breaking these degeneracies requires coordinated multi-frequency coverage.
	
	A related modeling uncertainty is the long-lived engine fraction. If only a fraction of BNS mergers produce a long-lived magnetar capable of sustained injection, the ejecta-channel detection rate scales linearly with that fraction. Our framework is explicitly constructed so that such population factors can be folded into the final rate estimates without altering the detectability maps.
	
	\subsection{Jet Structure and Inclination Selection Effects}
	\label{subsec:inclination}
	
	In our rate calculations, we adopt a conservative on-axis rate for a top-hat jet using a simple beaming factor $f_b \simeq 1-\cos\theta_j$. However, the actual detection yield of the jet channel may be modified by two competing effects. First, gravitational-wave selection effects naturally bias detected BNS mergers toward smaller inclination angles, effectively increasing the fraction of events with favorable viewing geometry relative to an isotropic sample. Second, if jets are ``structured"—where energy decreases smoothly with angle—the window for detecting the early bright peak expands to modestly off-axis observers.  While these factors may increase the jet detection probability, the jet's fleeting nature remains the primary operational hurdle. Even with an increased fraction of favorably oriented events, the detectability window for the jet component is limited to the first few hours or days post-merger. Consequently, the jet rate is suppressed by geometry despite higher brightness, reinforcing the need for rapid Tier-1 response.

	\subsection{Frequency Prioritization and the Role of Band 8}
	\label{subsec:band8}
	
	The 230 GHz band (Band 6) provides the largest detectable volume and the longest monitoring window for the isotropic ejecta channel. The 345 GHz band (Band~7) is best used as a follow-up diagnostic for events already detected or likely detectable at 230~GHz. In our fiducial ejecta models, the 230--460~GHz bands generally lie above both $\nu_a$ and $\nu_m$ near the light-curve peak, so the 230--345~GHz slope mainly tests the optically thin sub-mm spectrum and helps constrain the combination of shock energy, ambient density, and microphysical parameters. If a denser environment or earlier epoch shifts the turnover toward the mm/sub-mm bands, departures from the optically thin slope would provide evidence for spectral curvature. Band~8 (460~GHz) is therefore best regarded as a high-frequency consistency check for very nearby or unusually bright events, and as a potentially valuable early-time jet diagnostic under excellent observing conditions.

\subsection{Multiwavelength Synergies: FAST, WFST, and Beyond}
	\label{subsec:synergy}
	
	Sub-mm data are most powerful when integrated into a broader multiwavelength framework. For XSMT, coordination with domestic facilities like the Five-hundred-meter Aperture Spherical radio Telescope (FAST; \citealt{Nan2011}) for low-frequency radio monitoring and the Wide Field Survey Telescope (WFST; \citealt{Wang2023}) for early optical kilonova characterization is essential. FAST can anchor the low-frequency spectral slope, while XSMT captures the sub-mm peak and high-frequency spectral evolution. Joint modeling across these bands can distinguish between engine-driven outflows and purely kinetic-energy-driven shocks, providing a complete picture of the merger's caloric budget. The proposed cadence enables robust constraints on the merger environment and the physics of the relativistic outflows when paired with radio-to-X-ray data.

\subsection{\texorpdfstring{Source Confusion and Counterpart Identification}{Source Confusion and Counterpart Identification}}
\label{subsec:confusion}

A practical concern for any single-dish follow-up program is the confident association of a detected sub-mm source with the GW trigger. We assess this through three complementary lines of argument.

\textit{Chance coincidence.} The FoV/HPBW distinction introduced in Section~\ref{subsec:xsmt_thresholds} sets the relevant confusion problem. For a representative GW localization area of $\sim 10$--100~deg$^{2}$ in the LVK O4/O5 era \citep{Abbott2020ObsScenarios}, blind areal coverage would require hundreds to thousands of XSMT continuum-camera pointings, and the same region would contain $\sim 10^{5}$--$10^{7}$ independent beam areas. Sub-mm source counts from blank-field surveys (e.g., SCUBA-2, ALMA) give cumulative surface densities that can be high enough that a blind tiling of a full GW localization region would contain many unrelated field sources. Thus, XSMT follow-up should not be interpreted as an areal blind search. Instead, the relevant false-association probability is that for a targeted beam centered on a host-galaxy candidate or an independently localized EM counterpart. In that targeted mode, a static field source can be rejected by pre-existing imaging, repeat observations, or the absence of the expected week-to-month evolution. The primary practical background is therefore not the raw number of blank-field sub-mm sources in the full GW area, but unrelated transients or variables within the already-localized target field.

\textit{Temporal and spectral discrimination.} A BNS merger afterglow possesses characteristic features that distinguish it from common contaminants. AGN variability typically follows a red-noise power spectrum inconsistent with the smooth rise and power-law decay of a synchrotron blast wave. Stellar flares exhibit blackbody spectra peaking in the optical/NIR, unlike the non-thermal synchrotron spectrum ($F_\nu\propto\nu^{\alpha}$ with $\alpha\approx -0.65$ for $p=2.3$ above the SSA turnover) expected from a merger afterglow. Core-collapse supernovae can produce radio afterglows, but their sub-mm light curves are typically faster-evolving, and they are associated with star-forming host galaxies that can be identified in pre-existing optical imaging or promptly via WFST.

\textit{Multi-band confirmation.} Coordinated multi-band follow-up provides the cleanest counterpart check. A simultaneous detection at 230~GHz and 345~GHz with XSMT tests the sub-mm spectral slope and, when curvature is present, constrains the SSA turnover, providing a consistency check with the synchrotron afterglow model. A contemporaneous low-frequency counterpart (e.g., FAST at 1.4~GHz) anchors the lower-frequency spectrum, while an optical/near-IR non-detection (WFST) disfavors a thermal transient origin. In practice, a source that (i)~rises on week-to-month timescales, (ii)~displays a non-thermal SED consistent with a synchrotron afterglow, and (iii)~is positionally coincident with a GW-triggered host galaxy candidate, can be associated with the merger with less ambiguity.

\subsection{Operational Limitations and Future Improvements}
	\label{subsec:limitations}
	
	Our forecasts rely on fixed 1-hour $5\sigma$ thresholds, which assume stable weather and rapid scheduling. In practice, the detection probability is a convolution of the peak time $t_{\rm pk}$ and the observatory's cadence. If an event peaks during a period of poor weather or when the source is below the horizon, the effective detection rate will decrease. Future iterations of this framework could benefit from marginalized rate predictions that fold in astrophysical priors for $n_{\rm ext}$ and $M_{\rm ej}$, alongside realistic Monte Carlo simulations of observatory scheduling. The numerical pipeline developed here is designed for high reproducibility, allowing these practical constraints to be layered onto the model as XSMT nears full operations.

	\section{Conclusions}
	\label{sec:conclusions}
	
	This study provides a facility-specific forecast for the sub-mm detectability of GW counterparts using the upcoming XSMT. By coupling engine-dependent dynamics to a self-consistent synchrotron radiation model that accounts for SSA and DN effects, we have established a quantitative basis for XSMT's ToO program. Our primary findings are summarized as follows:
		
	\begin{enumerate}
		\item \textbf{The Isotropic Ejecta Channel as the Primary Discovery Route and Engine Diagnostic.}
			For a fiducial local event at $D=40$~Mpc, the isotropic ejecta afterglow represents the most reliable channel for characterization. It peaks on timescales of weeks to months and remains detectable for months to years at 230~GHz in standard 1-hour integrations. This long visibility window leaves room for delayed or repeated visits. A direct comparison with a BH-powered (injection-free) counterpart shows that the magnetar-boosted signal is brighter by a factor of $\sim 8\times 10^{3}$ (Section~\ref{subsec:fid_engine_comparison}), placing the BH-only case below the XSMT threshold at all times for a source at 40~Mpc. For the fiducial magnetar case, the corresponding expected all-sky detection rate is $\dot N_{\rm ej}\sim 0.05$--1.7~yr$^{-1}$ at 230~GHz based on current BNS merger-rate density, scaling linearly with the merger rate and with the long-lived engine fraction $f_{\rm mag}$.
		
		\item \textbf{Bimodal Light Curves and the Fleeting Jet Signal.} 
		The relativistic jet component, while intrinsically much brighter (peaking at $t \sim 10^{-2}$ days), provides only a fleeting window of hours for observation. Furthermore, its scientific yield is heavily throttled by geometric beaming factors. These bimodal characteristics—a short-lived jet peak followed by a long-lived ejecta peak—dictate the logic of our proposed observing cadence.
		
		\item \textbf{Strategic Frequency Prioritization.} 
			The 230~GHz band is the main discovery band for XSMT, offering the largest detectable volume for 1-hour integrations. The 345~GHz band provides the spectral information needed to test the optically thin sub-mm slope and to identify possible curvature toward the SSA turnover when paired with 230~GHz measurements. Band 8 (460~GHz) is best reserved for very nearby events, unusually bright ejecta afterglows, or early Tier-1 jet chasing under optimal atmospheric conditions.
		
		\item \textbf{An Optimized Two-Tier ToO Strategy.} 
		To maximize science return, we recommend a two-tier program: Tier-1 focuses on rapid jet chasing ($\lesssim 6$--12~h post-merger) for high-priority triggers with small localization volumes; Tier-2 employs a sparse long-baseline cadence (30--300+ days) optimized for the rising ejecta afterglow. This ``gap-aware" logic minimizes wasted resources while capturing the full evolution of the outflow.
		
		\item \textbf{The Dominance of Environmental Uncertainty.} 
		The circumburst density distribution $n_{\rm ext}$ remains the leading systematic uncertainty in sub-mm forecasts, as it directly acts as a gatekeeper for peak flux and timing. Coordinated follow-up with XSMT and other multi-wavelength facilities (e.g., FAST, WFST) is essential to break the microphysical degeneracies inherent in afterglow modeling.
	\end{enumerate}
	
			In summary, XSMT's dedicated ToO capability and the FoV/HPBW considerations discussed above make it best suited to targeted follow-up after prior GW or EM localization, and it is uniquely positioned to bridge the gap between optical kilonovae and late-time radio afterglows. These results provide a quantitative basis for allocating XSMT ToO time, prioritizing GW triggers by distance and multiwavelength indicators, and designing a follow-up cadence that can efficiently probe both early jet emission and late-time ejecta afterglows. Future work will extend these forecasts by folding astrophysical priors on $(M_{\rm ej}, n_{\rm ext})$, engine outcomes, and jet angular structure into a marginalized detection-rate calculation, and by incorporating cadence-dependent detection probability and weather/scheduling constraints. Such extensions will further refine the expected yield of XSMT in the multi-messenger era.
	
	\begin{acknowledgments}
		This work is supported by the National Key R\&D Program of China (2023YFA1608204), the National Natural Science Foundation of China (Grant  Nos. 12373052, 12321003, 12393813).
	\end{acknowledgments}

	\bibliography{xsmt_submm}
	\bibliographystyle{aasjournalv7}
	
	\appendix
	\renewcommand{\theHequation}{\thesection.\arabic{equation}}

    \section{Shared External-Shock Conventions and Kinematics}
    \label{app:shared_shock}
    
    This appendix summarizes the kinematic relations common to both the isotropic ejecta afterglow and the relativistic jet afterglow. By centralizing these shared relations, we ensure dynamical consistency across different outflow components and avoid repetition in the subsequent sections.  
    
    \subsection{Time and Redshift Convention}
    We evolve the dynamics using the source-frame arrival-time variable $t$. For observations at a given redshift $z$, the source-frame time is related to the observer-frame time $t_{\rm obs}$ by $t = t_{\rm obs}/(1+z)$. Similarly, the emitted frequency $\nu_{\rm em}$ and observed frequency $\nu_{\rm obs}$ follow the standard relation $\nu_{\rm em} = (1+z)\nu_{\rm obs}$. All spectral break frequencies calculated in our model are compared directly to $\nu_{\rm em}$ to determine the flux density.  
    
    \subsection{Radius Evolution in the Arrival-Time Formalism}
    The expansion of the shock front for both the ejecta and jet components is integrated according to the arrival-time radius evolution:
    \begin{equation}
    	\frac{dR}{dt}=\frac{\beta c}{1-\beta},\qquad
    	\beta=\sqrt{1-\Gamma^{-2}},
    	\label{eq:shared_dRdt}
    \end{equation}
    where $\Gamma$ is the bulk Lorentz factor. This formalism provides a monotonic mapping between the shock radius and observer time, which is essential for constructing EATS in our radiation calculations.  
    
    \subsection{Swept-up Mass and Solid Angle}
    We track the mass entrained from the circumburst medium using an explicit solid angle $\Omega(t)$. The total swept-up mass is given by:
    \begin{equation}
    	M_{\rm sw}(t)=\frac{\Omega(t)}{3}\,R^3\,n_{\rm ext}m_p,
    	\label{eq:shared_Msw}
    \end{equation}
    where $n_{\rm ext}$ is the ambient number density and $m_p$ is the proton mass. In our baseline treatment, the rate of mass accumulation is:
    \begin{equation}
    	\frac{dM_{\rm sw}}{dt}=\Omega(t)\,R^2\,n_{\rm ext}m_p\,\frac{dR}{dt}.
    	\label{eq:shared_dMswdt}
    \end{equation}
    For the isotropic ejecta, we adopt a constant $\Omega=4\pi$. For a top-hat jet, the solid angle is $\Omega(t)=2\pi\left[1-\cos\theta_j(t)\right]$, where the half-opening angle $\theta_j$ may evolve over time if lateral spreading is enabled.  
    
    \subsection{Comoving Age and Radiative Cooling}
    To account for radiative cooling at each shock radius, we compute the comoving age $t'$ of the emitting region. The evolution of comoving time is coupled to the expansion radius via:
    \begin{equation}
    	dt'=\frac{dR}{c\,\Gamma\,\beta}.
    	\label{eq:shared_tprime}
    \end{equation}
    This $t'$ is used to determine the cooling Lorentz factor $\gamma_c$, ensuring that the synchrotron spectrum accurately reflects the transition between the slow- and fast-cooling regimes in the sub-mm bands. 
    
    \section{Ejecta Dynamics with Engine Injection and Diffusion}
    \label{app:ejecta_dyn}
    
    This appendix supplies the diffusion, volume-expansion, and thermodynamic closure terms used by the energy-coupling equations and ejecta acceleration equation in Section~\ref{subsec:ejecta_dyn}.
    
    \subsection{Optical Depth and Energy Loss}
    As the ejecta expands, the photons trapped inside diffuse out, leading to energy loss. We track the optical depth $\tau$ using a gray opacity $\kappa$:
    \begin{equation}
    	\tau \simeq \kappa\,\rho'\,\frac{R}{\Gamma},\qquad
    	\rho'=\frac{M_{\rm ej}}{V'},
    \end{equation}
    where $\rho'$ is the comoving mass density and $V'$ is the comoving volume of the shell. The rate at which energy escapes the ejecta (the diffusion luminosity) is then estimated as \citep{Kasen2010}:
    \begin{equation}
    	L'_e \simeq \frac{E'_{\rm int}\,c}{(R/\Gamma)\,\max(\tau,1)},
    	\label{eq:Le_diff}
    \end{equation}
    where $E'_{\rm int}$ is the comoving internal energy. This term represents the ``cooling" of the shell due to thermal radiation. 

    The adiabatic-work term in Eq.~(\ref{eq:dEint_main}) is evaluated using the comoving volume-expansion rate of the spherical ejecta shell,
    \begin{equation}
    	\frac{dV'}{dt'} = 4\pi R^{2}\beta c,
    	\label{eq:dVprime_dtprime}
    \end{equation}
    so that $P'dV'/dt'$ gives the expansion work in the internal-energy equation.
    
    \subsection{Evolution of the Expansion Speed}
    For completeness, the expansion speed is obtained from Eq.~(\ref{eq:dGdt_main}) in Section~\ref{subsec:ejecta_dyn}. In that equation, $dE/dt$ is supplied by Eq.~(\ref{eq:dEdt_main}), $dE'_{\rm int}/dt'$ by Eq.~(\ref{eq:dEint_main}), $dV'/dt'$ by Eq.~(\ref{eq:dVprime_dtprime}), and $dM_{\rm sw}/dt$ by Eq.~(\ref{eq:shared_dMswdt}) with $\Omega=4\pi$ for the isotropic ejecta. The numerator of Eq.~(\ref{eq:dGdt_main}) represents the net power available to change the bulk kinetic motion after subtracting the internal-energy work term and the kinetic cost of sweeping up external material. The denominator is the effective inertia of the system, consisting of the ejecta rest-mass energy, comoving internal energy, and swept-up ISM contribution. This formulation ensures that the ejecta decelerates as it sweeps up the ISM, or accelerates if the engine injection term $\xi L_{\rm eng}(t)$ is sufficiently strong.
    
    
    \section{Jet Afterglow: Dynamics and Geometric Effects}
    \label{app:jet}
    
    In this appendix, we describe the equations specific to the relativistic jet component. While the general motion follows the same kinematic rules as the ejecta (see Appendix~\ref{app:shared_shock}), the jet's narrow shape and directional nature require a more detailed treatment of its energy and viewing geometry.  
    
    \subsection{Jet Energy and Geometry}
    
    We adopt a ``top-hat" model for the jet, meaning the energy is distributed uniformly within an initial half-opening angle $\theta_{j,0}$. Following the convention for a two-sided outflow, the \emph{true kinetic energy} in one side of the jet is defined as:
    \begin{equation}
    	E_{\rm jet} = E_{\rm iso}\,\frac{1-\cos\theta_{j,0}}{2},
    	\label{eq:Ejet_modeB}
    \end{equation}
    where $E_{\rm iso}$ is the isotropic-equivalent energy. The total swept solid angle at any time $t$ is:
    \begin{equation}
    	\Omega(t)=2\pi\left[1-\cos\theta_j(t)\right].
    	\label{eq:Omega_cone}
    \end{equation}
    To integrate the motion, we define an effective mass parameter $M_{\rm jet}$ based on the jet's initial energy and speed:
    \begin{equation}
    	M_{\rm jet}=\frac{E_{\rm jet}}{(\Gamma_0-1)c^2}.
    	\label{eq:Mjet_def}
    \end{equation}  
    
    \subsection{Speed and Deceleration}
    The Lorentz factor $\Gamma$ of the jet decreases as it sweeps up matter from the environment. Its evolution is tracked using the arrival-time variable $t$ via:
    \begin{equation}
    	\frac{d\Gamma}{dt}
    	=-\frac{\Gamma^2-1}{M_{\rm jet}+2\Gamma M_{\rm sw}}\frac{dM_{\rm sw}}{dt},
    	\label{eq:jet_dGdt}
    \end{equation}
    where $M_{\rm sw}$ is the mass collected within the jet's solid angle $\Omega(t)$. We use numerical sub-stepping in our code to maintain stability during the critical ``jet-break" phase, when the jet begins to slow down significantly.  
    
    \subsection{Sideways Spreading}
    As the jet slows, it can begin to expand sideways. We calculate the local sound speed $c_s$ in the comoving frame using a standard approximation for the gas pressure:
    \begin{equation}
    	\hat\gamma(\Gamma)\simeq\frac{4\Gamma+1}{3\Gamma},\qquad
    	c_s=c\left[\frac{\hat\gamma(\hat\gamma-1)(\Gamma-1)}{1+\hat\gamma(\Gamma-1)}\right]^{1/2}.
    	\label{eq:cs_code}
    \end{equation}
    The opening angle $\theta_j$ then grows at a rate determined by this sound speed:
    \begin{equation}
    	\frac{d\theta_j}{dt}=\frac{c_s}{\Gamma R(1-\beta)}\,\Theta_{\rm causal},
    	\label{eq:dthetadt}
    \end{equation}
    where $\Theta_{\rm causal}$ ensures that spreading only starts once the jet is ``causally connected" across its width (typically when $\Gamma\theta_j \le 1$). For cases requiring a more rapid expansion, our ``aggressive" mode adds an extra term to $dM_{\rm sw}/dt$ to account for the matter swept up by this sideways motion:
     \begin{equation}
    	\frac{dM_{\rm sw}}{dt}
    	=\Omega R^2 n_{\rm ext}m_p\frac{dR}{dt}
    	+\frac{1}{3}\frac{d\Omega}{dt}R^3 n_{\rm ext}m_p,
    	\label{eq:jet_dMswdt_aggressive}
    \end{equation}
    with
    \begin{equation}
    	\frac{d\Omega}{dt}=\frac{d\Omega}{d\theta_j}\frac{d\theta_j}{dt},\qquad
    	\frac{d\Omega}{d\theta_j}=2\pi\sin\theta_j.
    	\label{eq:dOmega_dtheta}
    \end{equation}
    
    \subsection{Equal-Arrival-Time Surfaces (EATS)}
    Because the jet is moving at near-light speeds, light from the edges of the jet reaches the observer later than light from the center. We account for this by integrating the emission over a surface where all light arrives at the same observer time $t_{\rm target}$:
    \begin{equation}
    	t_{\rm target}=t_{\rm on}(R;\text{patch})+\left(1-\cos\Psi\right)\frac{R}{c},
    	\label{eq:eats_condition}
    \end{equation}
    where $\Psi$ is the angle between the patch's velocity and the observer's line of sight. This geometric correction is vital for correctly predicting the early-time sub-mm peak and the subsequent decay.  

	\section{Synchrotron Spectrum with SSA and Deep-Newtonian Corrections}
	\label{app:ssa_dn}
	
	This appendix details the synchrotron emission model used for both the ejecta and jet components. Our implementation accounts for the transition from relativistic to non-relativistic regimes and the impact of self-absorption, ensuring robust flux predictions in the sub-mm bands. All spectral frequencies are defined in the emitted frame as $\nu_{\rm em}=(1+z)\nu_{\rm obs}$.  
	
	\subsection{Minimum Electron Lorentz Factor and Deep-Newtonian Handling}
	For an electron power-law index $p>2$, the theoretical minimum Lorentz factor is given by:
	\begin{equation}
		\gamma_{m,\rm raw} = 1 + \frac{p-2}{p-1} \frac{m_p}{m_e} \epsilon_e (\Gamma-1),
		\label{eq:gm_raw}
	\end{equation}
	where $\epsilon_e$ is the fraction of shock energy assigned to electrons. As the shock decelerates into the DN regime, $\gamma_{m,\rm raw}$ can approach unity \citep{Sironi2013}. To maintain physical consistency, we enforce a numerical floor $\gamma_{\rm DN}=2$:
	\begin{equation}
		\gamma_m = \max(\gamma_{m,\rm raw},\,\gamma_{\rm DN}).
		\label{eq:gm_clip}
	\end{equation}
	To conserve the total energy assigned to the electron population, we adjust the effective fraction of accelerated electrons, $f_{\rm acc}$, using a clipping function:
	\begin{equation}
		f_{\rm acc} = \mathrm{clip} \left( \frac{\max(\gamma_{m,\rm raw}-1,\,0)}{\max(\gamma_m-1,\,\varepsilon)} ,\, 0,\, 1 \right),
		\label{eq:facc}
	\end{equation}
	where $\varepsilon$ is a small numerical safety constant. The total number of radiating electrons is then $N_{e,\rm eff}=f_{\rm acc}N_e$. 
	
	\subsection{Magnetic Field and Characteristic Frequencies}
	We compute the comoving magnetic field $B'$ from the post-shock energy density using a formulation that remains valid throughout the mildly relativistic transition:
	\begin{equation}
		B' = \left[ 8\pi \epsilon_B (4\Gamma+3) (\Gamma-1) n_{\rm ext} m_p c^2 \right]^{1/2}.
		\label{eq:Bprime}
	\end{equation}
	The characteristic synchrotron frequency in the comoving frame, $\nu'_m$, and its boosted counterpart in the emitted frame, $\nu_m$, are calculated as:
	\begin{equation}
		\nu'_m = \frac{3}{4\pi} \frac{q_e B'}{m_e c} \gamma_m^2, \qquad \nu_m = \Gamma \nu'_m.
		\label{eq:num_prime_boost}
	\end{equation}  
	
	\subsection{SSA: Two-Branch Approximation}
	To determine the SSA frequency $\nu_a$, we adopt a practical two-branch approximation that accounts for whether self-absorption occurs above or below the characteristic frequency $\nu_m$ \citep{Wu2003}. Define a geometric parameter
	\begin{equation}
		\mathcal{B}\equiv
		\frac{q_e\,n_{\rm ext}\,R}{B'\,\gamma_m^5},
		\label{eq:ssa_base}
	\end{equation}
	and two coefficients (Euler Gamma function $\Gamma_{\rm E}$):
	\begin{align}
		C_1 &= \frac{16\sqrt[3]{4}\pi^2}{3\,\Gamma_{\rm E}(1/3)}\frac{p-1}{3}\frac{p+2}{3p+2},\\
		C_2 &= \frac{2^{3+p/2}\pi}{3\sqrt{3}}\frac{p-1}{3}\,
		\Gamma_{\rm E}\!\left(\frac{1}{6}+\frac{p}{4}\right)\,
		\Gamma_{\rm E}\!\left(\frac{11}{6}+\frac{p}{4}\right).
	\end{align}
	The two candidate SSA frequencies in the comoving frame are
	\begin{align}
		\nu'_{a,1} &= \left(C_1\mathcal{B}\right)^{3/5}\nu'_m,\\
		\nu'_{a,2} &= \left(C_2\mathcal{B}\right)^{2/(p+4)}\nu'_m.
	\end{align}
	The final comoving SSA frequency $\nu'_a$ is selected based on the ordering of the branches:
	\begin{equation}
		\nu'_a=
		\begin{cases}
			\nu'_{a,1}, & \nu'_{a,1}<\nu'_m,\\
			\nu'_{a,2}, & \nu'_{a,1}\ge \nu'_m,
		\end{cases}
		\label{eq:nua_branch}
	\end{equation}
	The emitted-frame frequency is $\nu_a = \Gamma \nu'_a$. This logic ensures that the sub-mm spectrum correctly captures the ``blocked" (optically thick) state at early times or in high-density environments.  
	
	\subsection{Peak Spectral Power and Flux Normalization}
	The observed brightness of the counterpart is determined by the peak specific power per electron in the comoving frame:
	\begin{equation}
		P'_{\nu',\max}=\frac{m_ec^2\sigma_T}{3q_e}\,B',
	\end{equation}
	which is boosted to the emitted frame as $P_{\nu,\max}=\Gamma P'_{\nu',\max}$. For a spherical shock, the total number of swept electrons is $N_e = \frac{4\pi}{3} R^3 n_{\rm ext}$. The peak flux density in the emitted frame is then normalized as:
	\begin{equation}
		F_{\nu,\max} = \frac{(1+z)\,N_{e,\rm eff}\,P_{\nu,\max}}{4\pi D_L^2},
	\end{equation}
	where the $(1+z)$ factor accounts for the standard relation between luminosity distance and per-Hz flux density.  
	
	\subsection{SSA-Modified Spectral Shape}
	The spectral profile is determined by the relative ordering of $\nu'_a$ and $\nu'_m$ \citep{Granot2002}. We define two regimes to describe the transition from optically thick to thin emission as below.
	
	\paragraph{Case A }($\nu'_a < \nu'_m$): The spectrum follows the standard sequence: $F_\nu \propto \nu^2$ (optically thick), $\nu^{1/3}$ (optically thin, below $\nu_m$), and $\nu^{-(p-1)/2}$ (optically thin, above $\nu_m$):
	\begin{equation}
		f_A(\nu')=
		\begin{cases}
			\left(\frac{\nu'_a}{\nu'_m}\right)^{1/3}\left(\frac{\nu'}{\nu'_a}\right)^2, & \nu'<\nu'_a,\\
			\left(\frac{\nu'}{\nu'_m}\right)^{1/3}, & \nu'_a\le \nu'\le \nu'_m,\\
			\left(\frac{\nu'}{\nu'_m}\right)^{-(p-1)/2}, & \nu'>\nu'_m.
		\end{cases}
	\end{equation}  
	
	\paragraph{Case B} ($\nu'_a \ge \nu'_m$): In this regime, self-absorption extends to higher frequencies, modifying the spectral shape:
	\begin{equation}
		f_B(\nu')=
		\begin{cases}
			\left(\frac{\nu'_m}{\nu'_a}\right)^{(p+4)/2}\left(\frac{\nu'}{\nu'_m}\right)^2, & \nu'\le \nu'_m,\\
			\left(\frac{\nu'_a}{\nu'_m}\right)^{-(p-1)/2}\left(\frac{\nu'}{\nu'_a}\right)^{5/2}, & \nu'_m<\nu'\le \nu'_a,\\
			\left(\frac{\nu'}{\nu'_m}\right)^{-(p-1)/2}, & \nu'>\nu'_a.
		\end{cases}
	\end{equation}  
	The final flux density at a given observer time and frequency is $F_\nu(\nu_{\rm obs},t_{\rm obs}) = F_{\nu,\max}(t) f(\nu')$.  
	
	\subsection{Late-Time Deep-Newtonian Phase}
	The deep-Newtonian phase begins at very late times when $\gamma_m-1\sim1$, The characteristic temporal decay index becomes \citep{Granot2006,Sironi2013}:
	\begin{equation}
		F_\nu(t_{\rm obs})\propto t_{\rm obs}^{\alpha_{\rm DN}},\qquad
		\alpha_{\rm DN}=-\frac{3(1+p)}{10}.
	\end{equation}
	In our baseline analysis, this override is restricted to the optically thin regime to maintain physical continuity.

	\section{Black-Hole Engine Option}
	\label{app:bh_engine}
	
	While the magnetar engine provides a powerful source of sustained energy, many BNS mergers are expected to result in the prompt formation of a BH. In this appendix, we describe the BH-engine baseline used to represent the ``prompt-collapse" scenario, which serves as a crucial reference for comparing against engine-driven models.  
	
	\subsection{Baseline: Prompt Collapse}
	In our standard BH scenario, we assume that no long-term energy injection occurs from the central remnant:
	\begin{equation}
		L_{\rm eng}(t)=0.
	\end{equation}
	In this case, the dynamics of the external shock are powered solely by the initial kinetic energy of the merger ejecta as it interacts with the circumburst medium. This ``no-injection" model defines the lower limit of sub-mm detectability for XSMT, relying entirely on the initial momentum of the outflow and internal heating from r-process decay. 
	
	\subsection{Exploratory Case: Fallback Accretion}
	For a more comprehensive analysis, we also include a phenomenological model to explore the effects of fallback accretion. Although not required for our primary conclusions, this allows us to test how a BH engine might still supply energy to the ejecta through a disk-driven outflow. We parameterize this additional injection as a power-law decay:
	\begin{equation}
		L_{\rm fb}(t)=L_{\rm fb,0},\max\left(1,\frac{t}{t_0}\right)^{-\alpha},
		\label{eq:bh_fallback}
	\end{equation}
	where $L_{\rm fb,0}$, $t_0$, and $\alpha$ are treated as free parameters. This approach provides a flexible way to study how sustained accretion modifies the peak flux and visibility window without requiring a specific, complex model of disk physics. By comparing this case against the magnetar scenario, we can better quantify the ability of sub-mm observations to distinguish between different types of central remnants.  
	
\end{document}